 \documentclass[preprint,12pt]{elsarticle}
\usepackage{multicol,amssymb,amsthm,amsmath,graphicx,geometry,float,times}
\usepackage{graphicx}
\usepackage{amsmath}
\usepackage{amssymb}
\usepackage{amsthm}
\usepackage{comment}
\newtheorem{theorem}{Theorem}[section]
\newtheorem{lemma}{Lemma}[section]
\newtheorem{remark}{Remark}[section]
\newtheorem{definition}{Definition}[section]
\newtheorem{observation}{Observation}[section]
\newtheorem{assumption}{Assumption}[section]
\numberwithin{equation}{section}
\journal{Journal of ...}

\begin{document}
\begin{frontmatter}
\title{Complete maximum likelihood estimation for SEIR epidemic models: theoretical development}
\author{Divine Wanduku$^{a*}$\footnote{a* Corresponding author.
} and Chinmoy Rahul$^{b}$ }
\address{$^{a}$Department of Mathematical Sciences,
Georgia Southern University, 65 Georgia Ave, Room 3042, Statesboro,
Georgia, 30460, U.S.A. E-mail:dwanduku@georgiasouthern.edu;wandukudivine@yahoo.com\\
$^{b}$Department of Mathematics and Statistics,University of Calgary,2500 University Drive NW,
Calgary, AB, Canada,
T2N 1N4. Email: chinmoyroy.rahul@ucalgary.ca, cr06998@georgiasouthern.edu}
\begin{abstract}
We present a class of  SEIR Markov chain models for infectious diseases observed over discrete time in a random human population living in a closed environment. The population changes over time through random births, deaths, and transitions between states of the population. The SEIR models consist of random dynamical equations for each state (S, E, I and R) involving driving events for the process. We characterize some special types of SEIR Markov chain models in the class including: (1) when birth and death are zero or non-zero, and (2) when the incubation and infectious periods are constant or random.  A detailed parameter estimation applying the maximum likelihood estimation technique and expectation maximization algorithm are presented for this study. Numerical simulation results are given to validate the epidemic models.
\end{abstract}

\begin{keyword}
 Discrete time Markov-chain\sep Chain-binomial models \sep transition events \sep birth-and-death subprocesses\sep MLE-technique\sep EM-algorithm\sep $m^{th}$ step MLE

\MSC 92B15 
\end{keyword}
\end{frontmatter}
%
%
	
	\section{INTRODUCTION}\label{chp1}
Some earliest well-known mathematical infectious disease models are deterministic. For example, the Kermack and McKendrick model\cite{kermack} is a SIR ordinary differential equation model. For more examples, see Hethcote\cite{hethcote-determ}. In fact, deterministic models play an important role, as first approximations to reality, to understand and identify underlying epidemiological factors controlling the eradication or persistence of diseases (cf. \cite{anderson}), and  estimating important epidemiologic parameters such as the basic reproduction number\cite{diekmann} etc. However, as nature is inevitably random over time, so do population events fluctuate over time, leading to a stochastic behavior of infectious disease dynamics. Thus, deterministic models leading to a single path for the disease dynamics, represent only the mean disease dynamics, while the stochastic analogs with multi-path representations offer a better approximation.

Compartmental mathematical models play an important role to investigate infectious disease epidemic dynamics. For instance,  influenza, malaria and other infectious disease deterministic models are studied in \cite{fe, mic, wan1}, whereas stochastic models based on diffusion processes are also utilized to study infectious diseases in \cite{yae, wan2, wanduku-malaria}. Pneumonia is studied in \cite{oj, emo, ams, gto, kel}.
     	In general, these compartmental epidemic models are classified as SIRS, SIR, SIS, SEIR and, SEIRS etc. models depending on the compartments of the disease states involved in the disease dynamics \cite{yae, md, tuckwell, fe, wan2, wan3}. Several authors devote interest to SEIR models \cite{mic, wan1,md,wanduku-malaria} which allow the inclusion of the exposed compartment (i.e. infected but not infectious), and lead to insights about the disease dynamics during the incubation period of the disease.

Probabilistic models also have a long history, for instance Bernoulli\cite{bernoulli}. In addition, stochastic epidemic models have been extensively studied (cf. \cite{Islam, andersson}). Modeling  with counting processes such as continuous-time Markov chains (CTMC) have wide applications in the literature\cite{bailey1975,allen-linda1, keeling-1}. In these models, the state of the process is integer valued and counts the number of susceptible, exposed, infectious or removed individuals(compartments) in the population over continuous time intervals. Discrete-time Markov chain (DTMC) epidemic models on the other hand, have also  received attention\cite{abbey,mgw,yae,tuckwell}. A usual assumption with DTMC model formulations is that the discrete time step is infinitesimally small such that only one transition at a time occurs between the disease states or compartments of the model, while multiple transitions occur with CTMC models\cite{allen-linda1}. Thus,  DTMC models  approximate the CTMC models, with more simplified transition probabilities over time, making their calculations and analysis less challenging for dynamic optimization and statistical estimation of system parameters
\cite{ross-tb, german,yae,keeling-1}.

A special class of DTMC epidemic models are chain-binomial epidemic models, and classical examples of these are the Greenwood\cite{mgw} and Reed frost\cite{abbey} models. These models are  called chain-binomial  models because their transition probabilities follow the binomial distribution. Some applications and complex extensions of these models have been studied\cite{tuckwell, yae, tsutsui, jgd}.

Estimating the parameters of a compartmental mathematical epidemic model serves as an important prelude to more accurate predictions about the epidemiological outcomes, and consequently formulating more rational data-informed public health policies. For instance, the basic reproduction number, denoted $R_{0}$, defined as the average number of secondary infectious cases that result from one infectious person placed in a complete disease-free population, is a complex parameter that depends on several other sub-epidemiological and demographical parameters of the infectious disease dynamic system. The true information about these sub-parameters are driven by data from the infectious disease dynamic system. Therefore, there is need to statistically infer these sub-parameters from the data, and consequently  obtain more informative estimates for $R_{0}$.

 There has been significant progress deriving and employing statistical and data-science  techniques to estimate and infer parameters of compartmental epidemic models, given data containing observations from the epidemic. A cross-section of some of these techniques  explored on either deterministic or stochastic compartmental epidemic models are given in the following \cite{ross-tb,german,yae,keeling-1,canto,alkema, riley-1,choib, yang-w,chritopzim, Gchowell,fierro-victoria}.

A special interest in this study is finding the maximum likelihood (ML)  estimators of some parameters for compartmental SEIR epidemic models suitable for describing the stochastic dynamics of diseases such as pneumonia and influenza etc. over discrete time intervals. The statistical ideas for ML estimation for some DTMC epidemic models have been explored in \cite{yae,chritopzim,fierro-victoria}, and CTMC epidemic models\cite{ross-tb,keeling-1}.

The method of ML estimation employed in infectious disease dynamic systems, seeks to find estimates for a set of epidemiological and demographical parameters from a given set of observations from the disease dynamic system, such  that, the estimates would maximize the chance of observing the given data from a population containing the parameters (cf. \cite{cb}). This estimation technique becomes challenging to apply, whenever minimizing the likelihood function leads to intractable results.  In such circumstances, the expectation-maximization algorithm (EM-algorithm)\cite{mgp, jba} is applied, whenever incorporating missing information from the given data leads to a more tractable likelihood function. These are the primary subjects of this paper. That is, to derive an adequate DTMC general model for SEIR epidemics such as Pneumonia or influenza epidemics etc. and to further explore the maximum likelihood estimation and EM-algorithm techniques to find MLE's for the vital parameters of the epidemic model.
Fierro et.al.\cite{fierro-victoria} experienced such challenges applying the ML-technique, and without an explicit estimator for the parameters, they instead investigated the consistency of the implicit estimators.

The Greenwood and Reed-Frost chain-binomial models  consider generations of infections, and infectious individuals no longer participate in subsequent disease transmission in another generation. This assumption is limiting and suitable for  disease dynamics, where the disease suddenly outbreaks in a given time generation, dies out, and reoccur in another time generation.
Yaesoubi and Cohen \cite{yae} also proposed a generalized class of DTMC models for infectious diseases involving multiple disease state compartments. They consider a hypothetical infectious disease with a natural history, that can be completely summarized with multiple serial classes. They studied various techniques to obtain dynamic optimal policies for their epidemic models. Fierro\cite{raul-fierro} has also considered a class of  DTMC epidemic models, studying the asymptotic consistency  between the stochastic models and their deterministic counterparts.

%

Building upon the ideas of the above studies \cite{abbey, yae, mgw, raul-fierro}, a SEIR DTMC epidemic model is proposed for diseases such as pneumonia transmitted by the \textit{S. pneumoniae} bacteria, or influenza transmitted by a unique strain of the influenza virus. It is assumed that the epidemic can be observed for a reasonable time interval, and over predetermined discrete times, and the population size is sufficiently large to allow binomial approximations for the transition probabilities between the disease states. The disease state of  individuals at any discrete time is either susceptible, exposed, infectious, or recovered with naturally acquired immunity, which is strong enough to protect the recovered person from subsequent infections by the same strain of the \textit{S. pneumoniae} bacteria or influenza virus .

Unlike \cite{yae,raul-fierro}, the presented DTMC SEIR model allows a framework that incorporates all transition events between states of the population apart from births and deaths (i.e the events of becoming exposed, infectious, and recovered), and also incorporates all birth and death events using random walk processes. That is, the presented framework allows for a constant finite, and random variable finite total human population at any time step. Moreover, this study also provides full analysis of the SEIR Markov chain model including the cases of fixed and random incubation and infectious periods in the disease dynamics. Furthermore, the technique of maximum likelihood estimation is fully developed and applied to find estimators for vital parameters of the disease model. Moreover, the method of expectation maximization algorithm is derived for the model, and applied to find the maximum likelihood estimators for the parameters of the model.

The rest of this paper is organized as follows. In Section~\ref{chp2}, we describe and derive the general class of SEIR Markov chain epidemic models,  characterizing the birth and death processes, and other transition sub-processes  of the general SEIR Markov chain model. In Section~\ref{chp3}, we derive the transition probabilities and feasible regions for some special SEIR Markov chain models, and also validate the epidemic models. In Section~\ref{chp4}, we find maximum likelihood estimators for important parameters of the SEIR Markov chain models.

\section{DESCRIPTION AND DERIVATION OF THE GENERAL SEIR MARKOV CHAIN}\label{chp2}
In this section, we describe adequately the SEIR disease epidemic in the human population. For simplicity, we use pneumonia in this description without loss of generality of all possible SEIR infectious diseases that follow the design of the epidemic model in this study exhibited in Figure~\ref{Fig 1}. We present the discretization of time; the decomposition of the human population into different classes involved in the pneumonia epidemic. We also characterize the birth and death sub-processes over time, and derive the general SEIR Markov chain model.

\subsection{Description of the SEIR infectious disease epidemic process}\label{ch2.sec1}

We consider a human population of size $N(t_k)>0$ at time $t_k$ living in a natural closed environment, where the outbreak of pneumonia occurs. During the period of the epidemic, it is assumed that birth occurs, and people collectively die from natural and disease related causes. Since the environment is closed, no migration occurs.

People vulnerable to infection who are not yet infected are denoted by $S$, and called the susceptible class. People who have been infected, but not infectious are denoted by $E$, and called the exposed class. The incubation period of the disease is denoted by $T_1$.
The people who are infected and spread pneumonia are denoted by $I$, and called the infectious class. The infectious class is treated against pneumonia over the period denoted by $T_2$, and recover with naturally acquired immunity. In general, it is assumed that $T_{1}\leq T_{2}$ .

\begin{figure}[H]
	\centering
	\includegraphics[width=10cm]{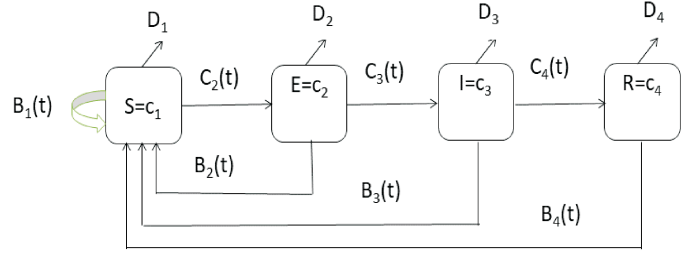}
	\caption{Shows the states of the system: $S, E, I, R$, and the transition between states
	 $C_i, i=1,2,3,4$, and also the births $B_i$, and deaths $D_i$ in the population.}
	\label{Fig 1}
\end{figure}

The naturally immune class is denoted by $R$. It is assumed that people who recover, acquire lifelong immunity against the strain of \textit{S. pneumoniae}.
It is also assumed that all births that occur are susceptible to pneumonia. A compartmental framework exhibiting the transition between the different states is given in Figure \ref{Fig 1}.

\subsection{Decomposition of the population over disease states and time}\label{ch2.sec2}
In this section,we characterize the different disease subclasses namely: susceptible, exposed, infectious and recovered individuals over discrete time intervals of fixed length, for example, hours, days, weeks, etc. The discretization process of time is presented in the following.
\begin{definition}{Time Discretization Process:}\label{ch2.sec2.def1}\par
	We use a regular partition $t_0, t_1, t_2,\ldots, t_k=t_0+(\Delta t)k, \forall k= 0,1,2,3,\ldots$ to create discrete time intervals of length $\Delta t$ ( $i.e.$ $[t_k, t_{k+1}),\forall k\ge 0$), and count the number of individuals of each compartment (susceptible, exposed, infectious or recovered class)  in each time interval. That is, the number of people in each state is counted over the sub-time intervals $[t_0,t_1), [t_1,t_2), [t_2,t_3),\ldots, [t_k,t_{k+1})$ $\forall k\ge 0$, where  $k \ge 0$ is a non negative integer. This time interval length $\Delta t$ is equivalent to a day, a week, a month etc.

	In this study, $x(t_k)$ represents the number of people in state $x \in \{S,E,I,R\}$ present at the beginning of the epoch $k$ ( $i.e.$ $[t_k, t_{k+1})$), or equivalently, at the end of the epoch $k-1$ ( $i.e.$ $[t_{k-1}, t_k)$). For example, $S(t_k)$ is the number of susceptible people present at the beginning of day $k$ ( $i.e.$ $[t_k, t_{k+1})$) or at the end of day $k-1$ ( $i.e.$ $[t_{k-1}, t_k)$). Thus, $S(t_k), E(t_k), I(t_k), R(t_k)\in \mathbb{Z}_{+}$, and $S(t_0)> 0, I(t_0)> 0$, $\forall k \in \mathbb{Z}_{+}$.
\end{definition}
\begin{definition}\label{ch2.sec2.def2}{Decomposition of the total population over time:}\par		
As discussed earlier, we subdivide the total population into four states: susceptible (S), exposed (E), infectious (I) and recovered (R). From Definition \ref{ch2.sec2.def1}, we define $N(t_k)$ to be the total human population present at the beginning of the epoch $k\ge 0$, or equivalently at the end of the epoch $(k-1), \forall k\ge 1$. Note that we synonymously use time $t_k$ and time $k$. These descriptions refer to the time characterization in Definition \ref{ch2.sec2.def1}. Furthermore, at time $k$, the total population present is given by
\begin{align}
N(t_k)= S(t_k)+ E(t_k)+ I(t_k)+ R(t_k)+ B(t_k)- D(t_k),
\end{align}
where $B(t_k)$ and $D(t_k)$ represent the total births and deaths, respectively, at time $k$.\par
In the absence of births and deaths, or when births and deaths are equal and cancel each other, the total population is given as follows:
	\begin{align}
	N(t_k) = S(t_k) + E(t_k) + I(t_k) + R(t_k),\quad and\quad N(t_k)=N(t_{k+1})\equiv N. \label{chp2.sec2.1.1.N(t_k)}
	\end{align}
	Also note that $N(t_k)\in \mathbb{Z}_+, \forall k\ge 1$,  $N(t_0)>0$, and $N\geq 0$ is a constant non-negative integer.
\end{definition}
\begin{definition}\label{ch2.sec2.def3}{Births and deaths over time:}\par
We consider birth and death in our model. We define $B(t_k)$ as the total birth in the population during the epoch $k$ ( $i.e$ $[t_k,t_{k+1})$). That is, $B(t_k)$ is the total birth count that occurs in the time interval $[t_k,t_{k+1})$, counted from the beginning of the interval $t_k$, until the onset $t_{k+1}$ of the next interval $[t_{k+1}, t_{k+1})$. Similarly, we define $B_x(t_k)$ as number of births by people of state $x \in \{S, E, I, R\}$ at time $k \ge 0$. It is easy to see that
\begin{align}
	B(t_k)= B_S(t_k) + B_E(t_k) + B_I(t_k) + B_R(t_k). \label{chp2.sec2.1.1.B(t_k)}
	\end{align}

 We also define $D(t_k)$ as total deaths that occur in the population during the epoch $k$ ( $i.e$ $[t_k,t_{k+1})$), where the death is counted over time interval $[t_k,t_{k+1})$, beginning at the point $t_k$ until the onset of the point $t_{k+1}$. Hence, $D(t_k)$ is the number of deaths at the end of the epoch $k$ or at the beginning of the epoch $k+1$. We similarly let $D_x(t_k)$ represent the deaths of state $x\in \{S, E, I, R\}$ at the epoch $k$. It is easy to see that
\begin{align}\label{chp2.sec2.1.1.D(t_k)}
D(t_k)= D_S(t_k)+ D_E(t_k)+ D_I(t_k)+ D_R(t_k).
\end{align}
Note that $B(t_k), B_x(t_k), D(t_k), D_x(t_k)\in \mathbb{Z}_+, x\in \{S, E, I, R\}$. Moreover, observe that $0 \le D_x(t_k)\le x(t_k), \forall x\in \{S, E, I, R\}$.  More elaboration of the birth and death processes over time, and some approximation of these processes in relation to the states $S(t_k), E(t_k), I(t_k), R(t_k)$ and $N(t_k)$ are given later.
\end{definition}
\begin{definition}\label{ch2.sec2.def4}{Transition events:}\par
We let $C_{ij}(t_k)$ be the number of epidemiological transition events that occur at time $k$, from state $i$ to state $j$, where $i,j\in \{S, E, I, R\}$. For example, $C_{SE}(t_k)$ represents the number of newly infected people becoming exposed during the time $[t_k,t_{k+1})$. $C_{EI}(t_k)$  and $C_{IR}(t_k)$  are the number of people converting from the exposed and infectious states,  to the infectious and removed states, respectively, during time interval $[t_k,t_{k+1})$.
 It follows that $C_{ij}(t_k)\in \mathbb{Z}_+, \forall i,j\in \{S, E, I, R\}$. Also, it is easy to see that
\begin{equation}
\left\{
\begin{array}{lllll}\label{chp2.sec2.1.2.transitionlimit}
0 &\le C_{SE}(t_k)+ D_{S}(t_k)\le S(t_k),\\
0 &\le C_{EI}(t_k)+ D_{E}(t_k)\le E(t_k),\\
0 &\le C_{IR}(t_k)+ D_{I}(t_k)\le I(t_k), \forall k\ge 0.\\
\end{array}
\right.
\end{equation}

Easily seen, at any time step $k>0$, the following important relationships can be deduced between the random variables: births $B(t_{k})$, deaths $D_{x}(t_{k}),\forall x\in \{S, E, I, R\}$ and transition events $C_{ij}(t_k),\forall i,j\in \{S, E, I, R\}$. (1) The random variables $B(t_{k})$ and $D_{x}(t_{k}),\forall x\in \{S, E, I, R\}$ are mutually independent, and the $D_{x}(t_{k})'s,\forall x\in \{S, E, I, R\}$ are also mutually independent. (2) $C_{SE}(t_k)+ D_{S}(t_k)$, $C_{EI}(t_k)+ D_{E}(t_k)$, and $C_{IR}(t_k)+ D_{I}(t_k)$ are mutually independent at time $k>0$. (3) It is possible that birth can occur in $C_{ij}(t_k)$, where $i,j\in \{S, E, I, R\}$. Thus, the $C_{ij}(t_k)'s$ and $B(t_{k})$ are mutually dependent as they measure mutually exclusive non-null events. Clearly, the pairs $(C_{SE}(t_k), D_{S}(t_k))$, $(C_{EI}(t_k), D_{E}(t_k))$, and $(C_{IR}(t_k), D_{I}(t_k))$ are dependent. These relationships will be useful to derive the transition probabilities of the stochastic process for the SEIR infectious disease epidemic.
\end{definition}

From Definition \ref{ch2.sec2.def1}- Definition \ref{ch2.sec2.def4}, it is easy to see that the susceptible, exposed, infectious and removal states of the population in the SEIR epidemic model at time $(k+1)$ $(S(t_{k+1}), E(t_{k+1}), I(t_{k+1}), R(t_{k+1}))$, given the states of the population at time $k$\\
 $(S(t_k), E(t_k), I(t_k), R(t_k))$, are guided by the following system of equations:
\begin{align}
S(t_{k+1})= S(t_k) + B(t_k) - D_S(t_k) - C_{SE}(t_k),\label{chp2.sec2.1.1.S(t_{k+1})}\\
E(t_{k+1})= E(t_k) - D_E(t_k) + C_{SE}(t_k) - C_{EI}(t_k),\label{chp2.sec2.1.1.E(t_{k+1})}\\
I(t_{k+1})= I(t_k) - D_I(t_k) + C_{EI}(t_k) - C_{IR}(t_k),\label{chp2.sec2.1.1.I(t_{k+1})} \\
R(t_{k+1})= R(t_k) - D_R(t_k) + C_{IR}(t_k).\label{chp2.sec2.1.1.R(t_{k+1})}
\end{align}
The following observations can be made from (\ref{chp2.sec2.1.1.S(t_{k+1})})- (\ref{chp2.sec2.1.1.R(t_{k+1})}).
\begin{observation}\label{chp2.sec2.1.1.obs1}
		\item[1] $N(t_{k+1})= N(t_k)+ B(t_k)- D(t_k), k=0,1,2,\ldots$. That is, the total population in the next time step changes only by birth and death from the total population of the previous time step.\label{chp2.ob1}
		\item[2] If B($t_k$) = D($t_k$) then N($t_{k+1}$) = N($t_k$), i.e. fixed population size at every time step.\label{chp2.ob2}
		\item[3] If B($t_k$) $\ge$ D($t_k$) then N($t_{k+1}$) $\ge$ N($t_k$), i.e. population is growing in size over time.\label{chp2.ob3}
		\item[4] If B($t_k$) $\le$ D($t_k$) then N($t_{k+1}$) $\le$ N($t_k$), i.e. population is decaying in size over time. We can imagine there is extinction of the population at some time $t_k$.\label{chp2.ob4}
\end{observation}

In the next section, we define a random process for the SEIR infectious disease epidemic process, characterize a feasible region for the stochastic process, and show that the stochastic process is a Markov chain.
\subsection{Derivation of the general SEIR population Markov chain}
Let $(\Omega, \mathfrak{F}, \mathbb{P})$ be a complete probability space and $\mathfrak{F}_{t_{k}}$ be a filtration (that is, sub $\sigma$- algebra $\mathfrak{F}_{t_{k}}$ that satisfies the following: given $t_{1}\leq t_{2} \Rightarrow \mathfrak{F}_{t_{1}}\subset \mathfrak{F}_{t_{2}}; E\in \mathfrak{F}_{t_{k}},\exists k$, and $P(E)=0 \Rightarrow E\in \mathfrak{F}_{t_{0}} $ ). Define a random vector measurable function
\begin{align}\label{chp2.sec2.4.feasible1}
X: \mathbb{Z}_+\times{\Omega} \rightarrow \mathbb{Z}^4_+,
\end{align}
where,
\begin{align}\label{chp2.sec2.4.feasible2}
X(t_k)=(S(t_k), E(t_k), I(t_k), R(t_k))\in \mathbb{Z}^4_+, \forall k \in \mathbb{Z}_+.
\end{align}
Moreover, the states $S(t_k), E(t_k), I(t_k)$, and $R(t_k)$ at any time $k\ge 0$, are $\mathfrak{F}_{t_{k}}$-measurable and satisfy the equations (\ref{chp2.sec2.1.1.S(t_{k+1})})- (\ref{chp2.sec2.1.1.R(t_{k+1})}). In addition, the random variables $C_{ij}(t_{k})\in \mathbb{Z}_+,\forall i,j\in \{S, E, I, R\}$, $B(t_k), B_x(t_k), D(t_k), D_x(t_k)\in \mathbb{Z}_+, x\in \{S, E, I, R\}$ are all discrete random variables on the probability space.
 The collection
\begin{align}\label{chp2.sec2.4.feasible3}
\{X(t_k): k\ge 0\}=\{(S(t_k), E(t_k), I(t_k), R(t_k)): k\ge 0\},
\end{align}
defines a random process for the SEIR infectious disease epidemic.

We derive the feasible region for the SEIR stochastic process $\{X(t_k): k\ge 0\}$ for the state of the process at time $k+1$ , given the state of the system at time $k,\forall k\geq 0$.
%
Solving the system (\ref{chp2.sec2.1.1.S(t_{k+1})})-(\ref{chp2.sec2.1.1.R(t_{k+1})}) for the transition events $C_{ij}(t_{k})\in \mathbb{Z}_+,\forall ij\in \{S, E, I, R\}$,  we have,
\begin{align}\label{chp2.sec.2.1.3.C_{SE}}
C_{SE}(t_k)= S(t_k)-S(t_{k+1}) + B(t_k) - D_S(t_k),
\end{align}
\begin{align}
C_{EI}(t_k)= E(t_k)-E(t_{k+1})+S(t_k)-S(t_{k+1})+ B(t_k)- D_S(t_k) - D_E(t_k),\label{chp2.sec.2.1.3.C_{EI}}
\end{align}
 and
\begin{align}
C_{IR}(t_k)=& I(t_k)-I(t_{k+1})+S(t_k)-S(t_{k+1})+E(t_k)-E(t_{k+1}) \nonumber\\
+& B(t_k)- D_S(t_k) - D_E(t_k) - D_I(t_k).\label{chp2.sec2.1.3.C_{IR}}
\end{align}
Also note that $C_{IR}$ from (\ref{chp2.sec2.1.1.R(t_{k+1})}) is given by,
\begin{align}
C_{IR}(t_k)=R(t_{k+1})-R(t_k)+D_R(t_k)\label{chp2.sec2.1.3.C_{IR}re}.
\end{align}
In addition, note that (\ref{chp2.sec2.1.1.R(t_{k+1})}) is only useful when $D_R(t_k)>0$, and redundant whenever $D_R(t_k)=0$.
If we introduce the notations
\begin{align}\label{chp2.sec2.4.2.relation}
& C_{12}(t_k)\equiv C_{SE}(t_k) , C_{23}(t_k)\equiv C_{EI}(t_k), C_{34}(t_k)\equiv C_{IR}(t_k),
S(t_k)\equiv X_1(t_k), \nonumber\\
&S(t_{k+1})\equiv X_1(t_{k+1}), E(t_k)\equiv X_2(t_k), E(t_{k+1})\equiv X_2(t_{k+1}), I(t_k)\equiv X_3(t_k), \nonumber\\
& I(t_{k+1})\equiv X_3(t_{k+1}), R(t_k)\equiv X_4(t_k), R(t_{k+1})\equiv X_4(t_{k+1}), D_S(t_k)\equiv D_1(t_k), \nonumber\\
& D_E(t_k)\equiv D_2(t_k),D_I(t_k)\equiv D_3(t_k),D_R(t_k)\equiv D_4(t_k),
\end{align}
then the transition events from (\ref{chp2.sec.2.1.3.C_{SE}})- (\ref{chp2.sec2.1.3.C_{IR}re}) can be written as
\begin{align}\label{chp2.sec2.3.4.tr}
C_{i,i+1}(t_k)+\sum_{j=1}^{i}D_j(t_k)-B(t_k)= \sum_{j=1}^{i}(X_j(t_k)-X_j(t_{k+1})),
\end{align}
and $C_{34}(t_k) - D_4(t_k) = X_4(t_{k+1})- X_4(t_k) $,
where, $i \in 1,2,3$.\par

Given the state $X(t_{k})$ of the process $\{X(t_k): k\ge 0\}$ at time $k$, we find restrictions for the transition events on the left hand side of (\ref{chp2.sec2.3.4.tr}),  in order to define a feasible region for the state $X(t_{k+1})$ at time $k+1$. Indeed, from (\ref{chp2.sec.2.1.3.C_{SE}}), it is easy to see from Definition \ref{ch2.sec2.def1}- Definition \ref{ch2.sec2.def4} and (\ref{chp2.sec2.1.2.transitionlimit}), that
\begin{align}\label{chp2.sec.2.4.3.C_{SE}1}
C_{SE}(t_k)+ D_S(t_k)- B(t_k)= S(t_k)-S(t_{k+1})\le S(t_k),
\end{align}
whenever birth and death are nonzero.

From (\ref{chp2.sec.2.1.3.C_{EI}}) and (\ref{chp2.sec2.1.2.transitionlimit}), it is easy to see that
\begin{align}\label{chp2.sec.2.4.3.C_{EI}1}
C_{EI}(t_k)+ D_S(t_k)+ D_E(t_k)- B(t_k)= E(t_k)-E(t_{k+1})+ S(t_k)-S(t_{k+1}),
\end{align}
and for nonzero birth and death,
\begin{align}\label{chp2.sec.2.4.3.ES}
E(t_k)-E(t_{k+1})+ S(t_k)-S(t_{k+1})\le E(t_k)+ S(t_k).
\end{align}

Similarly, from (\ref{chp2.sec2.1.3.C_{IR}}) and (\ref{chp2.sec2.1.2.transitionlimit}), it is easy to see that
\begin{align}
C_{IR}(t_k)+ D_S(t_k) + D_E(t_k) + D_I(t_k)- B(t_k)=& I(t_k)-I(t_{k+1})+ E(t_k)-E(t_{k+1})\nonumber\\
&+ S(t_k)-S(t_{k+1}),\label{chp2.sec2.1.3.C_{IR1}}
\end{align}
and for nonzero birth and death,
\begin{align}
I(t_k)-I(t_{k+1})+ E(t_k)-E(t_{k+1})+ S(t_k)-S(t_{k+1}) \le I(t_k)+ E(t_k)+ S(t_k).
\end{align}
Also, from (\ref{chp2.sec2.1.3.C_{IR}re}), it is easy to see that
\begin{align}
C_{IR}(t_k)- D_R(t_k)= R(t_{k+1})-R(t_k)\le I(t_k)\label{chp2.sec2.1.3.C_{IR}relim}.
\end{align}

From (\ref{chp2.sec2.3.4.tr})- (\ref{chp2.sec2.1.3.C_{IR}relim}) we now define the feasible region for the state $X(t_{k+1})$ of the process $\{X(t_k), k\ge 0\}$, whenever birth and death are zero or non-zero, and  given the state $X(t_{k})$ of the process is known.
\begin{theorem}\label{chp2.sec2.4.feasible1.thm1}
Let the assumptions in Definition \ref{ch2.sec2.def1}- Definition \ref{ch2.sec2.def4} hold, and the stochastic process $\{X(t_k): k\ge 0\}$ defined in (\ref{chp2.sec2.4.feasible1})- (\ref{chp2.sec2.4.feasible3}) satisfy the system of equations (\ref{chp2.sec2.1.1.S(t_{k+1})}) - (\ref{chp2.sec2.1.1.R(t_{k+1})}). Then the following hold:
\begin{description}
\item[1.] When birth and death processes are zero, the feasible region $\Omega^{1}_{X(t_{k+1})}$ for the state $X(t_{k+1})$ of the process at time $k+1$, given the state of the process $X(t_k)$ at time $k$, is defined as follows:
\begin{align}\label{chp2.sec2.4.feasible4}
\Omega^{1}_{X(t_{k+1})}=& \{(s_{k+1}, e_{k+1}, i_{k+1}, r_{k+1})\in \mathbb{Z}^4_+|0\le S(t_k) - s_{k+1}\le S(t_k), \nonumber\\
&0\le E(t_k)- e_{k+1}+ S(t_k) - s_{k+1} \le E(t_k), \nonumber\\
&0\le I(t_k)- i_{k+1}+ E(t_k)- e_{k+1}+ S(t_k) - s_{k+1}\le I(t_k),\nonumber\\
\quad and \quad & 0\le r_{k+1}- R(t_k)\le I(t_k)\},
\end{align}
and using the notations in (\ref{chp2.sec2.4.2.relation}),
\begin{align}\label{chp2.sec2.4.feasible5}
\Omega^{1}_{X(t_{k+1})}=\{(X_1,X_2,X_3,X_4)\in\mathbb{Z}^4_{+}|& 0\le \sum_{j=1}^{i}(X_j(t_k)-X_j)\le X_i(t_k),\nonumber\\
&0\le X_4 - X_4(t_k)\le X_3(t_k)\}, i=1,2,3.
\end{align}
\item[2.] When birth and death processes are nonzero, the feasible region $\Omega^{2}_{X(t_{k+1})}$ of the process at time $k+1$, given the state at time $k$, is defined as follows:
\begin{align}\label{chp2.sec2.4.feasible6}
\Omega^{2}_{X(t_{k+1})}&= \{(s_{k+1}, e_{k+1}, i_{k+1}, r_{k+1})\in \mathbb{Z}^4_+|0\le S(t_k) - s_{k+1}\le S(t_k), \nonumber\\
&0\le E(t_k)- e_{k+1}+ S(t_k) - s_{k+1} \le E(t_k)+ S(t_k),0\le I(t_k)- i_{k+1} \nonumber\\
&+ E(t_k)- e_{k+1}+ S(t_k) - s_{k+1}\le I(t_k)+ E(t_k)+ S(t_k),\nonumber\\
\quad and \quad & 0\le r_{k+1}- R(t_k)\le I(t_k)\},
\end{align}
and using the notations in (\ref{chp2.sec2.4.2.relation}),
\begin{align}\label{chp2.sec2.4.feasible7}
\Omega^{2}_{X(t_{k+1})}=\{(X_1,X_2,X_3,X_4)\in\mathbb{Z}^4_{+}|& 0\le \sum_{j=1}^{i}(X_j(t_k)-X_j)\le \sum_{j=1}^{i}X_j(t_k),\nonumber\\
&0\le X_4 - X_4(t_k)\le X_3(t_k)\}, i=1,2,3.
\end{align}
\end{description}
\end{theorem}
\begin{proof}
See (\ref{chp2.sec2.3.4.tr}) - (\ref{chp2.sec2.1.3.C_{IR}relim}).
\end{proof}
Observe that $\Omega^{1}_{X(t_{k+1})}\subset \Omega^{2}_{X(t_{k+1})}$. This signifies that the occurrence of birth and death in the population expands the state space of the SEIR Markov chain model $\{X(t_k): k\ge 0\}$. Also observe that when birth and death are zero, that is, $B(t_{k})=D(t_{k})=0$, the reduced vector $X(t_{k})=(S(t_k), E(t_k), I(t_k)), k\geq 0$ is sufficient to describe the SEIR model (\ref{chp2.sec2.1.1.S(t_{k+1})})-(\ref{chp2.sec2.1.1.R(t_{k+1})}), since (\ref{chp2.sec2.1.3.C_{IR}re}) becomes redundant.

\begin{remark}
	Suppose the disease dynamics consists of $M$ serial disease states $X_1, X_2,\ldots,X_M$ structured with births- $B_1, B_2,\ldots,B_M$, deaths- $D_1,D_2,\ldots,D_M$, and transition events\\
 $C_{12}, C_{23},\ldots,C_{i,i+1},\ldots,C_{M-1,M}$, with a similar design as in Figure~\ref{Fig 1}. Then the following generalization for each transition event $C_{i,{i+1}}(t_k), i=1,2,\ldots,M-1$, can be obtained,
	\begin{align}\label{chp2.sec.2.3.2cii}
	C_{i,i+1}(t_k) +\sum_{j=1}^{i}D_j(t_k) - B(t_k)= \sum_{j=1}^{i}(X_j(t_k)-X_j(t_{k+1})),
	\end{align}
	where, $B(t_k)= \sum_{j=1}^{M}B_j(t_k)$ and   $i \in 1,2,3,\ldots,(M-1)$. Or equivalently,
	\begin{align}
	C_{i,i+1}(t_k) - \sum_{j=i+1}^{M}D_j(t_k)= \sum_{j=i+1}^{M}(X_j(t_{k+1})-X_j(t_k)),
	\end{align}
	where, $i \in 1,2,3,...,(M-1)$.
\end{remark}

We introduce new notations in the following, in addition to (\ref{chp2.sec2.4.feasible1}) - (\ref{chp2.sec2.4.feasible3}). (i.) Let $x_k\in \mathbb{Z}^4_{+}$, where $x_k= (x^k_1,x^k_2,x^k_3,x^k_4)\equiv(s_k,e_k,i_k,r_k)\in \mathbb{Z}^4_{+}$. That is, $x^k_1\equiv s_k,x^k_2\equiv e_k, x^k_3\equiv i_k$ and $x^k_4\equiv r_k$.  The vector $x_k=(s_k,e_k,i_k,r_k)\in \mathbb{Z}^4_{+}$ consists of non-negative integers for each $k \in\{0,1,2,3,\ldots\}$, and $X(t_k)=x_k$ if and only if
\begin{align}
&S(t_k)\equiv X_1(t_k)=x^k_1, S(t_{k+1})\equiv X_1(t_{k+1})=x^{k+1}_1,
E(t_k)\equiv X_2(t_k)=x^k_2, \nonumber\\
&E(t_{k+1})\equiv X_2(t_{k+1})=x^{k+1}_2,
I(t_k)\equiv X_3(t_k)=x^k_3, I(t_{k+1})\equiv X_3(t_{k+1})=x^{k+1}_3,\nonumber\\
&R(t_k)\equiv X_4(t_k)=x^k_4,R(t_{k+1})\equiv X_4(t_{k+1})=x^{k+1}_4\label{chp2.sec.2.3.3stonot2},
\end{align}
where, $x^k_1\equiv s_k,x^k_2\equiv e_k, x^k_3\equiv i_k$ and $x^k_4\equiv r_k$. (ii.)The notation $G(t_{k})|H(t_{k})$ denotes a conditional random variable $G(t_{k})$ depending on the random variable $H(t_{k})$ at each time $k\geq 0$ in the usual way. That is, for each $k\geq 0$, given a value for  $H(t_{k})$, then $G(t_{k})$ is determined. Moreover, the collection $\{G(t_{k})|H(t_{k}), k\geq 0\}$ is called a sub-stochastic process of the process $\{H(t_{k}), k\geq 0\}$.

 The following result proves that $\{X(t_k): k\ge 0\}$ in $(\ref{chp2.sec2.4.feasible3})$ is a Markov chain.
 \begin{theorem}\label{chp2.sec.2.1.3.lemma1}
	The stochastic process $\{X(t_k): k=0,1,2,\ldots\}$ is  a discrete time Markov chain, and the transition probabilities are completely defined by the distribution of the random variables $C_{i,i+1}(t_k),i=1,2,3, B(t_k)$ and $D(t_k), \forall k\ge 0$. Moreover, the general form of the transition probabilities is given as follows.
\item[1.] If births and deaths are non-zero at every time step, that is, suppose the conditional random variables denoted $B(t_{k})|X(t_{k}),\forall k\geq 0$  and $ D_{x}(t_{k})|X(t_{k}),\forall k\geq 0, x\in \{S, E, I, R\} $ define sub-stochastic processes describing births and deaths in the population, respectively, where for each $X(t_{k})$, $B(t_{k})\geq 0$, $D_{x}(t_{k})\geq 0$, then
\begin{eqnarray}
&&P(X(t_{k+1})=x^{k+1}|X(t_k)=x^k)\nonumber\\
&&=\sum_{b^{k}=0}^{\infty}\sum^{x^{k}_{1}}_{d^{k}_{1}=0}\sum^{x^{k}_{2}}_{d^{k}_{2}=0}\sum^{x^{k}_{3}}_{d^{k}_{3}=0}\sum^{x^{k}_{4}}_{d^{k}_{4}=0}P(B(t_{k})=b^{k}|X(t_{k})=x^{k})
\prod^{4}_{j=1}P(D_{j}(t_{k})=d^{k}_{j}|X(t_{k})=x^{k})\nonumber\\
&&\prod_{i=1}^{3}P\left(C_{i,{i+1}}(t_k)
=b^{k}-\sum_{j=1}^{i}d^{k}_{j}+\sum_{j=1}^{i}(x^k_j-x^{k+1}_j)|B(t_{k})=b^{k},(D_{j}(t_{k})=d^{k}_{j})_{j=1}^{4},  X(t_k)=x^k;\right),\nonumber\\
\label{chp2.sec.2.1.3.lemma1.eq1}
\end{eqnarray}
where, $k \in\{0,1,2,...\}$.
\item[2.] If there are no births and deaths at every time step, that is, suppose the conditional random variables describing births and death in the population are zero, that is,  $B(t_{k})|X(t_{k})\equiv 0|X(t_{k}),\forall k\geq 0$  and $ D_{x}(t_{k})|X(t_{k})\equiv 0|X(t_{k}),\forall k\geq 0, x\in \{S, E, I, R\} $, respectively, then
	\begin{eqnarray}
&&P(X(t_{k+1})=x^{k+1}|X(t_k)=x^k)\nonumber\\
&&=\prod_{i=1}^{3}P\left(C_{i,{i+1}}(t_k)
	=\sum_{j=1}^{i}(x^k_j-x^{k+1}_j)|X(t_k)=x^k\right),\label{chp2.sec.2.1.3.lemma1.eq2}
\end{eqnarray}
where $k \in\{0,1,2,...\}$.
\end{theorem}
\begin{proof}
	We first show that $\{X(t_k):k=0,1,2,\ldots\}$ is a Markov chain. That is, we show that it satisfies the Markov property. In other words, we show that,
	\begin{align}\label{chp2.sec2.3.4rhs}
	RHS &\equiv P(X(t_{k+1})=x^{k+1}|X(t_k)=x^k,X(t_{k-1})=x^{k-1},\ldots,X(t_0)=x^0)\nonumber\\
	&= P(X(t_{k+1})=x^{k+1}|X(t_k)=x^k)\equiv LHS.
	\end{align}
	The RHS of (\ref{chp2.sec2.3.4rhs}) is written as follows:
	\begin{align}
	RHS& \equiv P(X(t_{k+1})=x^{k+1}|X(t_k)=x^k,X(t_{k-1})=x^{k-1},...,X(t_0)=x^0)\nonumber\\
	 &=P(X_1(t_{k+1})=x^{k+1}_1,X_2(t_{k+1})=x^{k+1}_2,X_3(t_{k+1})=x^{k+1}_3,X_4(t_{k+1})=x^{k+1}_4|\nonumber\\&(X_1(t_k)=x_1^k,X_2(t_k)=x_2^k,X_3(t_k)=x_3^k,X_4(t_k)=x_4^k),(X_1(t_{k-1})=x_1^{k-1},\nonumber\\&X_2(t_{k-1})=x_2^{k-1}, X_3(t_{k-1})=x_3^{k-1},X_4(t_{k-1})=x_4^{k-1}),\ldots,(X_1(0)=x^0_1,\nonumber\\&X_2(0)=x^0_2,X_3(0)=x^0_3,X_4(0)=x^0_4)).\label{chp2.sec2.3.4.xtt}
	\end{align}
	Using the expression in (\ref{chp2.sec2.3.4.tr}), the RHS is written as follows:
	\begin{align}
	RHS\equiv& P(C_{12}(t_k)=x^k_1-x^{k+1}_1+B(t_{k})-D_1(t_k),C_{23}(t_k)=x^k_2-x^{k+1}_2+x^k_1-x^{k+1}_1\nonumber\\&-D_1(t_k)-D_2(t_k)+B(t_k),C_{34}(t_k)=x^k_3-x^{k+1}_3+x^k_2-x^{k+1}_2+x^k_1-x^{k+1}_1\nonumber\\
&-D_1(t_k)-D_2(t_k)-D_3(t_k)+B(t_k), C_{34}(t_k)=x^{k+1}_4-x^k_4+D_4(t_k)|(X_1(t_k)=x_1^k,\nonumber\\&X_2(t_k)=x_2^k,X_3(t_k)=x_3^k,X_4(t_k)=x_4^k),(X_1(t_{k-1})=x_1^{k-1},X_2(t_{k-1})=x_2^{k-1},\nonumber\\
&X_3(t_{k-1})=x_3^{k-1},X_4(t_{k-1})=x_4^{k-1}),\ldots,(X_1(0)=x^0_1,X_2(0)=x^0_2,\nonumber\\&X_3(0)=x^0_3,X_4(0)=x^0_4))\label{chp.2.sec2.3.4.rlhs2}\\
	 &=P(C_{12}(t_k)=x^k_1-x^{k+1}_1+B(t_k)-D_1(t_k),C_{23}(t_k)=x^k_2-x^{k+1}_2+X^k_1\nonumber\\
&-x^{k+1}_1-D_1(t_k)-D_2(t_k)+B(t_k),C_{34}(t_k)=x^k_3-x^{k+1}_3+x^k_2-x^{k+1}_2+x^k_1\nonumber\\
&-x^{k+1}_1-D_1(t_k)-D_2(t_k)-D_3(t_k)+B(t_k), C_{34}(t_k)=x^{k+1}_4-x^k_4+D_4(t_k)|\nonumber\\
&(X_1(t_k)=x_1^k,X_2(t_k)=x_2^k,X_3(t_k)=x_3^k,X_4(t_k)=x_4^k)).\label{chp.2.sec2.3.4.rlhs}
	\end{align}
	Note that (\ref{chp.2.sec2.3.4.rlhs2}) reduces to (\ref{chp.2.sec2.3.4.rlhs}), since the driving events $C_{i,i+1}(t_k),i=1,2,3, B(t_k)$ and $D_{x}(t_{k})|X(t_{k}),\forall k\geq 0, x\in \{S, E, I, R\}$ at time $t_k$ depend only on the state $X(t_k)$. Also, applying the relationships between the random variables $C_{i,i+1}$'s, $(D^{k}_{j})_{j=1}^{4}$ and $B(t_{k})$ in Definition~\ref{ch2.sec2.def4}, and basic probability rules, the result in (\ref{chp2.sec.2.1.3.lemma1.eq1}) follows immediately. It follows trivially that setting the random variables   representing birth and death terms to zero, the result in (\ref{chp2.sec.2.1.3.lemma1.eq2}) also follows immediately.
%
\end{proof}

Observe from Theorem~\ref{chp2.sec.2.1.3.lemma1}(1.) that there are several possible discrete time and discrete state sub-stochastic processes $\{B(t_{k})|X(t_{k}), k\geq 0\}$  and $ \{D_{x}(t_{k})|X(t_{k}), k\geq 0\}, x\in \{S, E, I, R\} $ to represent the random births $B(t_{k})$ and deaths $D_{x}(t_{k})\forall x\in \{S, E, I, R\}$  over time $k\geq 0$, respectively, given $X(t_{k}$. To completely characterize the transition probability in (\ref{chp2.sec.2.1.3.lemma1.eq1}), we consider some examples of the sub-processes $\{B(t_{k})|X(t_{k}), k\geq 0\}$  and $ \{D_{x}(t_{k})|X(t_{k}), k\geq 0\}, x\in \{S, E, I, R\} $. 

\section{SOME SPECIAL SEIR INFECTIOUS DISEASE MARKOV CHAIN MODELS}\label{chp3}
In this section we consider some special SEIR infectious disease Markov chain models of the class of SEIR models $\{X(t_k): k\ge 0\}$  guided by (\ref{chp2.sec2.1.1.S(t_{k+1})}) - (\ref{chp2.sec2.1.1.R(t_{k+1})}), and defined in  Theorem~\ref{chp2.sec.2.1.3.lemma1}. The special cases are based on whether the total  population $N(t_{k}), \forall k\geq 0$ defined in Definition~\ref{ch2.sec2.def2} is a constant at each time $k\geq 0$ or a stochastic process. Recall Observation~\ref{chp2.ob1} states that the population size in the closed environment is fixed over time either in the absence of birth and death, or whenever birth and death are equal at each time step. Note that the use of the assumption of fixed total population size in this paper refers to the former.
\subsection{Birth and death sub-stochastic processes}\label{chp3.subsec1}
The stochastic process  $\{B(t_{k})|X(t_{k}), k\geq 0\}$ can be characterized for simplicity using a homogeneous Poisson process as follows. Supposes births occur independently and at a constant birthrate of $\lambda_{b}$ per unit time. Let $\tilde{B}(t_{k}),k\geq 0$ represent total births over $[t_{0},t_{k}]$, then $\tilde{B}(t_{k})$  can be formulated easily from Definition~\ref{ch2.sec2.def3} as follows
  \begin{equation}\label{chp3.subsec1.eq1}
    \tilde{B}(t_{k})=\sum_{i=0}^{k}B(t_{i})|X(t_{i}), \quad k\geq 0.
  \end{equation}
  Thus, the stochastic process $\{\tilde{B}(t_{k}),k\geq 0\}$ is a Poisson process with rate $\lambda_{b}$, defined as a random walk process in (\ref{chp3.subsec1.eq1}) with only births (pure birth process). Moreover,  the conditional random variable $B(t_{k})|X(t_{k})$ (in Definition~\ref{ch2.sec2.def3}) is indeed an increment of the Poisson process $\{\tilde{B}(t_{k}),k\geq 0\}$, and has Poisson distribution, with mean $\lambda_{b}(t_{k+1}-t_{k})=\lambda_{b}\Delta t$. Therefore, the stochastic process $\{B(t_{k})|X(t_{k}), k\geq 0\}$ is a collection of Poisson random variables over discrete time $k\geq 0$ with mean $\lambda_{b}\Delta t$.

  The sub-stochastic process $\{D_{x}(t_{k})|X(t_{k}), k\geq 0\}, x\in \{S, E, I, R\} $ can also be characterized using a homogenous Poisson process, and Binomial distribution. Suppose deaths  occur in the state $x\in \{S, E, I, R\}$ of the population independently and at a constant deathrate $\mu_{d_{x}}$ per unit time, then the number of deaths in the state $x\in \{S, E, I, R\}$ of the population over time follows homogenous Poisson process with mean $\mu_{d_{x}}$, and the random lifetime until death $T>0$ of an individual has exponential distribution with mean $\frac{1}{\mu_{d_{x}}}$ and survival probability denoted $\bar{S}_{x}(t)=e^{-\mu_{d_{x}} t}$. Thus, the probability that an individual at time $k>0$ will die
  \begin{equation}\label{chp3.subsec1.eq2}
    P_{d_{x}}(t_{k})=1-\frac{\bar{S}_{x}(t_{k+1})}{\bar{S}_{x}(t_{k})}=1-e^{-\mu_{d_{x}} \Delta t},\quad x\in \{S, E, I, R\} .
  \end{equation}
Since individuals of state $x\in \{S, E, I, R\}$ of the population die independently with probability in  $(\ref{chp3.subsec1.eq2})$,  the stochastic process $\{D_{x}(t_{k})|X(t_{k}), k\geq 0\}, x\in \{S, E, I, R\} $ is a collection of binomial random variables with parameters $Binomial(X(t_{k}), P_{d_{x}}(t_{k}) )$.

More generally, if the random lifetime until death $T>0$ is some other distribution with better failure rates, e.g. Weibull distribution, $W(a_{x}, b_{x})$, then it is easy to see using the formula in (\ref{chp3.subsec1.eq2}) that $P_{d_{x}}(t_{k})=1-e^{-\left[(a_{x}t_{k+1})^{b_{x}}-(a_{x}t_{k})^{b_{x}}\right] },\quad x\in \{S, E, I, R\}$ and the stochastic process $\{D_{x}(t_{k})|X(t_{k}), k\geq 0\}, x\in \{S, E, I, R\} $ is collection of binomial random variables with parameters
 $Binomial(X(t_{k}), P_{d_{x}}(t_{k}) )$.

 From the above, the birth and death related probability terms $P(B(t_{k})=b^{k}|X(t_{k})=x^{k})$, and $
P(D_{j}(t_{k})=d^{k}_{j}|X(t_{k})=x^{k})$ in the transition probability (\ref{chp2.sec.2.1.3.lemma1.eq1}) in Theorem~\ref{chp2.sec.2.1.3.lemma1}(1.) are  defined. To completely specify (\ref{chp2.sec.2.1.3.lemma1.eq1}), we now characterize the distribution of the conditional random variables $\left(C_{i,{i+1}}(t_k)
\vert B(t_{k}),(D_{j}(t_{k}))_{j=1}^{4},  X(t_k)\right)$, $\forall i\in \{1,2,3\}$, whenever the random variables $B(t_{k}),(D_{j}(t_{k}))_{j=1}^{4},  X(t_k)$ are given. That is, we characterize the sub-stochastic processes $\left\{\left(C_{i,{i+1}}(t_k)
\vert B(t_{k}),(D_{j}(t_{k}))_{j=1}^{4},  X(t_k)\right); k\geq 0\right\}$, $\forall i\in \{1,2,3\}$.

To optimize space for parameter estimation, we proceed with the  SEIR model with constant population size (i.e. $N(t_{k})=N$, $N>0$ constant), with general transition probabilities given in Theorem~\ref{chp2.sec.2.1.3.lemma1}(2.), i.e. whenever births and deaths are zero. The complete description of the SEIR model $\{X(t_k): k\ge 0\}$, in Theorem~\ref{chp2.sec.2.1.3.lemma1}(1.), whenever $N(t_{k}), \forall k\geq 0$ is a stochastic process, including the full characterization of the sub-stochastic processes $\left\{\left(C_{i,{i+1}}(t_k)
\vert B(t_{k}),(D_{j}(t_{k}))_{j=1}^{4},  X(t_k)\right); k\geq 0\right\}$, $\forall i\in \{1,2,3\}$, and the transition probability given by (\ref{chp2.sec.2.1.3.lemma1.eq1}) for nonzero births and deaths, will appear in \cite{wanduku-diffusion}. In the following we characterize the sub-stochastic processes $\{C_{i,{i+1}}(t_k)
| X(t_k), k\geq 0\}, \forall i\in \{1,2,3\}$ (i.e. births and deaths are zero).
\subsection{Transitional events sub-stochastic processes}
In the absence of births and deaths in the population, i.e. $N(t_{k})=N,\forall k\geq 0$, $N>0$ constant. The feasible region for the SEIR Markov chain model $\{X(t_k): k\ge 0\}$ in this case is given in (\ref{chp2.sec2.4.feasible4}). We describe the sub-stochastic processes $\{C_{i,{i+1}}(t_k)
| X(t_k), k\geq 0\}, \forall i\in \{1,2,3\}$ (i.e. births and deaths are zero) for two cases (1) when the incubation and infectious periods, $T_1$ and $T_2$, respectively, are constant and equal to $\Delta t$, and (2) when $T_1$ and $T_2$ are random variables with  lifetime distributions. The following assumptions are utilized. 
\begin{assumption}\label{chp2.sec2.assum2}
		\item[1.] Let $p$, be the probability of passing infection to a susceptible person after one interaction with an infectious person at any time $t_k$, $k\geq 0$.

		 \item[2.]
 Let the stochastic process $\{n(t_k),k\ge 0\}$ be a Poisson process with rate $\lambda$ representing the number of people a susceptible individual meets over time until the epoch $k$ $([t_k, t_{k+1}))$ (e.g. in week $k\geq 0$). That is, $ n(t_k)$ is the  total number of people a susceptible individual meets and interacts with over the total time interval $[t_{0}, t_{k+1})$,  $\forall k\geq 0$. 
		Thus, the increment $n(t_{k+1})- n(t_k)$, $\forall k\ge 0$ is the number of people the susceptible person interacts with during epoch $k$, where the epoch starts from the point $t_k$ until the onset of the point $t_{k+1}$. It is easy to see that
		\begin{align}\label{chp3.sec1.npois}
		n(t_{k+1})- n(t_k) \equiv n(\Delta t) \sim Poisson (\lambda \Delta t).
		\end{align}
	
		\item[3.] For each $k\ge 0$, let $Y^i_n (t_k)$ count the infectious people the $i^{th}$ susceptible person meets in the epoch $k$ $([t_k, t_{k+1}))$, given that $n(t_{k+1})-n(t_k) = n$ people were met during that epoch $k$, where $i=1,2,3,\ldots, S(t_k)$.	Under the assumptions of (1) independent contacts in the population, and (2) homogenous mixing so that all contacts are equally likely regardless of the state (susceptible, exposed, infectious or removed) of an individual in the population, then it is easy to see that
\begin{align}\label{chp3.sec1.yin}
		Y^i_n(t_k)\sim Binomial (n(t_{k+1})- n(t_k) = n, \alpha^i(t_k)),
		\end{align}
where $\alpha^i(t_k)$- the probability that the $i^{th}$ susceptible interacts with an infectious person in the population, given the  $N(t_k)$- total people present at time $k$ $([t_k, t_{k+1}))$ is given as follows:
		\begin{align}\label{chp2.sec1.alfa}
		\alpha^i(t_k)= \frac{I(t_k)}{N(t_k)-1}.
		\end{align}
		\item[4.] We let the categorical random variable $Z^i(t_k)$  indicate the  $i^{th}$ susceptible person getting infected at time $k$ $([t_k, t_{k+1}))$, and let $p^i(t_k)$ be the probability that the $i^{th}$ susceptible person gets infected at time $k$.

\end{assumption}
We utilize Assumption \ref{chp2.sec2.assum2}, $1-4$ to find the probability  $p^i(t_k)$, whenever the total population at time $k$, is a constant, i.e. $ N(t_k) = N>0$ is fixed.
\begin{theorem}\label{chp2.sec.2.1.3.lemma2}
Under the conditions of Assumption \ref{chp2.sec2.assum2} above, and also for $N(t_k) = N(t_{k+1}) =N>0$ constant (i.e. $B(t_k)=D(t_k)= 0$) in Observation \ref{chp2.sec2.1.1.obs1}, then the probability that a susceptible person gets infected at time $k$ $([t_k,t_{k+1}))$ is given as follows:
\begin{align}\label{p^i(t_k)}
p^i(t_k) = 1-e^{-p\alpha^i(t_k)\lambda \Delta t}.
\end{align}
\end{theorem}
\begin{proof}
	Applying the laws of probability,
	\begin{align}\label{chp2.lemma2.prof11}
	p^i(t_k) &= P(Z^i(t_k)=1|N(t_k)=N) \nonumber\\
	&= \sum_{n=0}^{\infty}\sum_{j=0}^{n} P(Z^i(t_k)=1, Y^i_n(t_k)=j,n(\Delta t)=n|N(t_k)=N),\nonumber\\
	&= \sum_{n=0}^{\infty}\sum_{j=0}^{n}	P(Z^i(t_k)=1|Y^i_n(t_k)=j,n(\Delta t)=n,N(t_k)=N)\times\nonumber\\
	&\times P(Y^i_n(t_k)=j|n(\Delta t)=n,N(t_k)=N).P(n(\Delta t)=n|N(t_k)=N).
	\end{align}
	By Assumption \ref{chp2.sec2.assum2} and similar reasoning in \cite{abbey}, it is easy to see that for each $n = 0,1,2,\ldots; j=0,1,2,\ldots,n$, and $k\ge 0$,
	\begin{align}
	&P(Z^i(t_k)=1|Y^i_n(t_k)=j,n(\Delta t)=n,N(t_k)=N)=[1-(1-p)^j].\label{chp2.lemma2.prof12}\\
	\quad Also, \quad\nonumber\\
	&P(Y^i_n(t_k)=j|n(\Delta t)=n,N(t_k)=N)=\binom{n}{j}(\alpha^i(t_k))^j(1-\alpha^i(t_k))^{n-j},\label{chp2.lemma2.prof13}\\
	\quad and \quad \nonumber\\
	&P(n(\Delta t)=n|N(t_k)=N)=\frac{e^{-\lambda\Delta t}(\lambda\Delta t)^n}{n!}.\label{chp2.lemma2.prof14}
	\end{align}
	Substituting (\ref{chp2.lemma2.prof12})-(\ref{chp2.lemma2.prof14}) into (\ref{chp2.lemma2.prof11}), we get,
	\begin{align}\label{chp2.lemma2.prof15}
	p^i(t_k) &= \sum_{n=0}^{\infty}\sum_{j=0}^{n}[1-(1-p)^j]\binom{n}{j}(\alpha^i(t_k))^j(1-\alpha^i(t_k))^{n-j}\frac{e^{-\lambda\Delta t}(\lambda\Delta t)^n}{n!}\nonumber\\
	&=1-\sum_{n=0}^{\infty}\sum_{j=0}^{n}(1-p)^j\binom{n}{j}(\alpha^i(t_k))^j(1-\alpha^i(t_k))^{n-j}\frac{e^{-\lambda\Delta t}(\lambda\Delta t)^n}{n!}.
	\end{align}
	Let $q=1-p$, then
	\begin{align}\label{chp2.lemma2.prof16}
	\sum_{j=0}^{n}(1-p)^j\binom{n}{j}(\alpha^i(t_k))^j(1-\alpha^i(t_k))^{n-j}&=\sum_{j=0}^{n}\binom{n}{j}(\alpha^i(t_k)q)^j(1-\alpha^i(t_k))^{n-j}\nonumber\\
	&=[\alpha^i(t_k)q+1-\alpha^i(t_k)]^n\nonumber\\
	&=[\alpha^i(t_k)(1-p)+1-\alpha^i(t_k)]^n\nonumber\\
	&=[1-\alpha^i(t_k)p]^n.
	\end{align}
	Substituting (\ref{chp2.lemma2.prof16}) into (\ref{chp2.lemma2.prof15}), we have
	\begin{align}
	p^i(t_k) = 1-\sum_{n=0}^{\infty}[1-\alpha^i(t_k)p]^n\frac{e^{-\lambda\Delta t}(\lambda\Delta t)^n}{n!}.\label{prob}
	\end{align}
	Again let $\beta(t_k)=1-\alpha^i(t_k)p$. Then,
	\begin{align}
	\sum_{n=0}^{\infty}[1-\alpha^i(t_k)p]^n\frac{e^{-\lambda\Delta t}(\lambda\Delta t)^n}{n!}&=\sum_{n=0}^{\infty}(\beta(t_k))^n\frac{e^{-\lambda\Delta t}(\lambda\vartriangle t)^n}{n!}\nonumber\\
	&=e^{-\lambda\Delta t}\sum_{n=0}^{\infty}\frac{(\lambda\Delta t\beta(t_k))^n}{n!}\nonumber\\
	&=e^{-\lambda\Delta t}e^{-\lambda\Delta t\beta(t_k)}\nonumber\\
	&=e^{-\lambda\Delta t(1-\beta(t_k))}\nonumber\\
	&=e^{-\lambda\Delta t\alpha^i(t_k)p}.\label{exp}
	\end{align}
	Substituting (\ref{exp}) into (\ref{prob}), we obtain (\ref{p^i(t_k)}).
%
	
\end{proof}
\begin{remark}\label{ch2.sec2.1.4.int1}
The probability that the $i^{th}$ susceptible person gets infected at time $k\ge 0$, i.e. $ p^i(t_k), k\ge 0$  in (\ref{p^i(t_k)}) in Theorem \ref{chp2.sec.2.1.3.lemma2} can be interpreted as follows. Observe from (\ref{p^i(t_k)}) that the term $ p\alpha^i(t_k)$ represents the probability of that the $i^{th}$ susceptible person meets and gets infection from one random infectious person at time $k$ $([t_k,t_{k+1}))$. Since the Poisson rate $\lambda$ is the average number of people (infectious or noninfectious) that the $i^{th}$ susceptible person meets per unit time, then assuming independent contacts per unit time, it follows that $ p\alpha^i(t_k)\lambda\Delta t$ is the binomial expected number of infectious people the $i^{th}$ susceptible person interacts with over an interval of length $\Delta t$, which results to infection of the susceptible person.

Therefore, suppose the conditional random variable $\hat{n}_{k}|n(t_{k+1})-(t_{k})$ is the Poisson random number of infectious people the $i^{th}$ susceptible person meets in the epoch $k$, (i.e. in the interval $[t_k, t_{k+1})$ of length $\Delta t$) who \underline{\textit{almost surely}}  infect the susceptible individual, then $\hat{n}_{k}|n(t_{k+1})-(t_{k})\sim Poisson (\mu = p\alpha^i(t_k)\lambda \Delta t)$.
 Moreover. it is easy to see that $T$- the random time until the $i^{th}$ susceptible person meets an infectious person who almost surely passes infection has an exponential distribution with mean $\frac{1}{\mu}$. Thus, utilizing the formula with survival distribution functions in (\ref{chp3.subsec1.eq2}), an alternative representation for $p^i(t_k)$ in (\ref{p^i(t_k)})  is the following:
 \begin{align}
 p^i(t_k)= 1-\frac{\bar{S}_{x}(t_{k+1})}{\bar{S}_{x}(t_{k})}=1-e^{-p\alpha^i(t_k)\lambda \Delta t}, x=S.\label{chp2.sec2.1.4.int}
 \end{align}
 \end{remark}
\subsection{Transition probabilities for the SEIR model with equal incubation and infectious periods}\label{chp2.sec.2.1.5}
Using Theorem~\ref{chp2.sec.2.1.3.lemma2}, we characterize the sub-stochastic processes $\{C_{i,{i+1}}(t_k)
| X(t_k), k\geq 0\}, \forall i\in \{1,2,3\}$ (i.e. births and deaths are zero) for the case where the incubation and infectious periods, $T_1$ and $T_2$, are constant and equal. Moreover, we completely derive the transition probabilities for the SEIR Markov chain model $\{X(t_k):k=0,1,2,\ldots\}$ defined in Theorem~\ref{chp2.sec.2.1.3.lemma1}[2.].

 The following assumptions are utilized. (1.) A newly infected person at time $t_{k}$ will be exposed for one time unit, after which the person becomes infectious by time $t_{k+1}$, i.e. $T_{1} =\Delta t$. (2.) It is assumed that all newly infectious individuals at the beginning of epoch $k$ $([t_k, t_{k+1}))$, will be identified and treated, or completely recovered from the disease by the beginning of epoch $k+1$. That is, the infectious period for every individual $T_2$ is given by $T_2 = \Delta t$.
\begin{theorem}\label{chp2.thm1}
	Let Theorem \ref{chp2.sec.2.1.3.lemma2} be satisfied. Under the assumption that $B(t_k)=D(t_k)=0$, $\forall k\geq 0$,  the SEIR Markov chain model $\{X(t_k):k=0,1,2,\ldots\}$ has the following transition probabilities
	 \begin{align}\label{chp3.thm3.2.tran1}%
	&P(X(t_{k+1})=x^{k+1}|X(t_k)=x^k)\nonumber\\
	&=\begin{cases}
	\binom{s_k}{s_{k+1}}(p^i(t_k))^{s_k-s_{k+1}}(1-p^i(t_k))^{s_{k+1}}, &\text{for $(s_{k+1},e_{k+1},i_{k+1})\in\Omega^{1}_X(t_{k+1}),$}\\
	0, &\text{otherwise,}
	\end{cases}
	\end{align}
	whenever the incubation period $T_1$ and infectious period $T_2$ are constant, and equal to one time unit $\Delta t$. Moreover, the feasible region for the chain in (\ref{chp2.sec2.4.feasible4}) reduces to
	\begin{align}\label{chp3.thm3.2.tran1.eq1}
	\Omega^{1}_X(t_{k+1})=&\{(s_{k+1},e_{k+1},i_{k+1})\in\mathbb{Z}^3_{+}\vert 0\le s_{k+1} \le s_k, 0\le e_{k+1} \le s_k, 0\le i_{k+1} \le s_k,\nonumber\\
	&s_{k+1}+e_{k+1}=s_k, i_{k+1}=e_k \}.
	\end{align}
\end{theorem}
	\begin{proof}
	From Theorem \ref{chp2.sec.2.1.3.lemma1}[2.], the general form of transition probabilities when $B(t_k) = D(t_k) = 0$ is given as,
	\begin{align}\label{chp3.sec3.1.2product}
	P(X(t_{k+1})|X(t_k)=x^k)=\prod_{i=1}^{3}P(C_{i,{i+1}}(t_k)=\sum_{j=1}^{i}(x^k_j-x^{k+1}_j)|X(t_k)=x^k),
	\end{align}
	where $k \in\{0,1,2,\ldots\}$.\par
	From equation (\ref{chp2.sec.2.1.3.C_{SE}}) we have that in the absence of birth and death,
	\begin{align}
	C_{SE}(t_k)\equiv C_{12}(t_k)= S(t_k)-S(t_{k+1}). \label{chp2.sec.2.1.5.C_{SE}2}
	\end{align}
	Similarly from equation (\ref{chp2.sec.2.1.3.C_{EI}}) it is easy to see that when $T_1 = \Delta t$, $C_{EI}(t_k)=E(t_k)$, and
	\begin{align}
	C_{EI}(t_k)\equiv C_{23}(t_k)= E(t_k)-E(t_{k+1})+S(t_k)-S(t_{k+1})=E(t_k),\label{chp2.sec.2.1.5.C_{EI}2}
	\end{align}
	Also from equation (\ref{chp2.sec2.1.3.C_{IR}}) observe that for $T_2 = \Delta t$, $C_{IR}(t_k)=I(t_k)$, and
	\begin{align}
	C_{IR}(t_k)\equiv  C_{34}(t_k)&= I(t_k)-I(t_{k+1})+S(t_k)-S(t_{k+1})+E(t_k)-E(t_{k+1})=I(t_k). \label{chp2.sec2.1.5.C_{IR}2}
	\end{align}
 From (\ref{chp2.sec.2.1.5.C_{SE}2})-(\ref{chp2.sec2.1.5.C_{IR}2}), we obtain
	\begin{align}
	I(t_{k+1}) = E(t_k). \label{chp2.sec.2.1.5.I(t_{k+1})}
	\end{align}
	From equation (\ref{chp2.sec.2.1.5.C_{SE}2})-(\ref{chp2.sec.2.1.5.I(t_{k+1})}) it is easy to see that
	\begin{align}\label{chp3.fesiblesk}
	0\le S(t_{k+1}) \le S(t_k), 0\le E(t_{k+1}) \le S(t_k) \quad and \quad 0\le I(t_{k+1}) \le S(t_k).
	\end{align}
	Thus, the feasible region from (\ref{chp2.sec.2.1.5.C_{SE}2}) - (\ref{chp3.fesiblesk}), and letting $S(t_k)=s_k, E(t_k)=e_k, I(t_k)=i_k, R(t_k)=r_k, \forall k\geq 0$ is defined as follows
	\begin{align}
	\Omega^{1}_X(t_{k+1})=&\{(s_{k+1},e_{k+1},i_{k+1})\in\mathbb{Z}^3_{+}\vert 0\le s_{k+1} \le s_k, 0\le e_{k+1} \le s_k, 0\le i_{k+1} \le s_k,\nonumber\\
	&s_{k+1}+e_{k+1}=s_k, i_{k+1}=e_k \}.
	\end{align}
	Recall Theorem {\ref{chp2.sec.2.1.3.lemma2}}, the probability, $p^i(t_k)$, that a susceptible person gets infected at time $k$ $([t_k,t_{k+1}))$, whenever $B(t_k) = D(t_k) = 0$  is defined in (\ref{p^i(t_k)}). Thus, it is easy to see from the conditions of Assumption~\ref{chp2.sec2.assum2} that the random number of new exposed persons converting from the susceptible class at time $k\geq 0$, $C_{SE}(t_k)$, has the binomial distribution $Binomial(S(t_k)=s_k, p^i(t_k))  $. That is,
\begin{align}\label{chp2.sec.2.1.5.C_SE}
&P(C_{SE}(t_k)=c^k_{SE}|X(t_k)=x^k)\nonumber\\
&=\begin{cases}
\binom{s_k}{c^k_{SE}}(p^i(t_k))^{c^k_{SE}}(1-p^i(t_k))^{s_k-c^k_{SE}}, & \text{for $c^k_{SE}=0,1,2,\ldots,s_k$},\\
0, &\text{otherwise.}
\end{cases}
\end{align}
But, from (\ref{chp2.sec.2.1.5.C_{SE}2}), writing (\ref{chp2.sec.2.1.5.C_SE}) in terms of  $S(t_k)=s_k, E(t_k)=e_k, I(t_k)=i_k, R(t_k)=r_k, \forall k\geq 0$, it is easy to see that
\begin{eqnarray}\label{chp2.sec.2.1.5.C_SE3}
&&P(C_{SE}(t_k)=c^k_{SE}|X(t_k)=x^k)= P(S(t_{k+1})=s_{k+1}|X(t_k)=x^k)\nonumber\\
&&=
\begin{cases}
\binom{s_k}{s_{k+1}}(p^i(t_k))^{s_k-s_{k+1}}(1-p^i(t_k))^{s_{k+1}}, & \text{for $s_{k+1}=0,1,2,\ldots,s_k$},\\
0, &\text{otherwise.}
\end{cases}
\end{eqnarray}
Since $T_1=T_2=\Delta t$,  $C_{EI}(t_k)=E(t_k)$, and  $C_{IR}(t_k)=I(t_k)$
 it is easy to see that
\begin{equation}\label{chp.2.sec.2.1.5.C_EI3}
  P(C_{EI}(t_k)=c^k_{EI}|X(t_k)=x^k)=
  \begin{cases}
  1, & \text{for $c^k_{EI}= e_k$},\\
  0, &\text{otherwise,}
  \end{cases}
  \end{equation}
  and 
    \begin{equation}\label{chp.2.sec.2.1.5.C_IR3}
  P(C_{IR}(t_k)=c^k_{IR}|X(t_k)=x^k)=
  \begin{cases}
  1, & \text{for $c^k_{IR}=i_k$},\\
  0, &\text{otherwise.}
  \end{cases}
  \end{equation}
  From (\ref{chp3.sec3.1.2product}), (\ref{chp2.sec.2.1.5.C_SE3}), (\ref{chp.2.sec.2.1.5.C_EI3}) and (\ref{chp.2.sec.2.1.5.C_IR3}) we have
  \begin{align}
  &P(X(t_{k+1})=x^{k+1}|X(t_k)=x^k)= P((S(t_{k+1}),E(t_{k+1}),I(t_{k+1})=(s_{k+1},e_{k+1},i_{k+1})|\nonumber\\
  &X(t_k)=x^k)= P(C_{SE}(t_k)=c^k_{SE}|X(t_k)=x^k)\times P(C_{EI}(t_k)=c^k_{EI}|X(t_k)=x^k)\times\nonumber\\
  &\times P(C_{IR}(t_k)=c^k_{IR}|X(t_k)=x^k),\nonumber\\
  &= \begin{cases}
  \binom{s_k}{s_{k+1}}(p^i(t_k))^{s_k-s_{k+1}}(1-p^i(t_k))^{s_{k+1}}, &\text{for $(s_{k+1},e_{k+1},i_{k+1})\in\Omega_X(t_{k+1})$},\\
  0, &\text{otherwise.}
  \end{cases}
  \end{align}	
\end{proof}	

\subsection{Transition probabilities for the SEIR model with random incubation and infectious periods }
Similarly, using Theorem~\ref{chp2.sec.2.1.3.lemma2}, we characterize the sub-stochastic processes $\{C_{i,{i+1}}(t_k)
| X(t_k), k\geq 0\}, \forall i\in \{1,2,3\}$ (i.e. births and deaths are zero) for the case where the incubation and infectious periods, $T_1$ and $T_2$, are random variables. Moreover, we completely derive the transition probabilities for the SEIR Markov chain model $\{X(t_k):k=0,1,2,\ldots\}$ defined in Theorem~\ref{chp2.sec.2.1.3.lemma1}[2.]. This scenario is guided by the following assumptions.

Note that various lifetime distributions can be used to represent the distributions of  $T_{1}$ and $T_{2}$. We consider a simply scenario where $T_{1}$ and $T_{2}$ are exponentially distributed.
Assume that individuals who are exposed become infectious independently and at a constant average rate of $\delta_e$ per unit time. Then $\{M_1(t_k), k=0,1,2,\ldots\}$ is a Poisson process, with rate $\delta_e$, where $M_1(t_k), \forall k\geq 0$ represents the number of people converting from the exposed into the infectious state over time interval $(0,t_k]$. Thus, it is easy to see that $T_1$, the time until an exposed person becomes infectious follows exponential distribution with mean $\frac{1}{\delta_e}$.

Similarly,  assume that individuals who are infectious recover independently, and at a constant average rate of $\delta_r$ per unit time. Thus, $\{M_2(t_k), k=0,1,2,\ldots\}$ is a Poisson process with rate $\delta_r$, where $M_2(t_k), \forall k\geq 0$ represents the number of people converting from the infectious state into the recovery state over time $(0,t_k]$. Therefore, it is easy to see that $T_2$, the time until an infectious person becomes recovered follows exponential distribution with mean $\frac{1}{\delta_r}$.

Using the survival distribution formula  (\ref{chp3.subsec1.eq2}), it is easy to see that the probabilities that the $i^{th}$ exposed  and infectious persons convert into the infectious and removed states, respectively, in the interval $[t_k,t_{k+1})$ are given as follows:
\begin{align}\label{chp.2.sec.2.1.7.ex.prob}
P^i_{EI}(t_k)= 1-\frac{\bar{S}_{E}(t_{k+1})}{\bar{S}_{E}(t_{k})}=1-P(M_1(t_{k+1})-M_1(t_k)=0)= 1- e^{-\delta _e \Delta t},\forall i=1, 2,\ldots, e_{k},
\end{align}
and
\begin{align}\label{chp.2.sec.2.1.7.inf.prob}
P^i_{IR}(t_k)= 1-\frac{\bar{S}_{I}(t_{k+1})}{\bar{S}_{I}(t_{k})}=1-P(M_2(t_{k+1})-M_2(t_k)=0)= 1- e^{-\delta _r \Delta t},\forall i=1, 2,\ldots, i_{k}.
\end{align}
%
Applying similar reasoning in Subsection~\ref{chp2.sec.2.1.5} we characterize the process $\{C_{i,{i+1}}(t_k)
| X(t_k), k\geq 0\}, \forall i\in \{1,2,3\}$ (i.e. births and deaths are zero), and completely derive the  transition probabilities for the SEIR Markov chain model $\{X(t_k), k=0,1,2,\ldots\}$ when the above conditions are satisfied. Due to limited space, we present the main results and further comments will appear in \cite{wanduku-diffusion}.
\begin{theorem}\label{chp.2.sec.2.1.7.inf.prob.thm1}
Suppose the assumptions of Theorem \ref{chp2.sec.2.1.3.lemma1}, and Theorem \ref{chp2.sec.2.1.3.lemma2} are satisfied. Also, let $B(t_k) = D(t_k) = 0$, and suppose conversions from the exposed and infectious states to the infectious and removal states, respectively, are described by independent Poisson processes $\{M_1(t_k), k=0,1,2,\ldots\}$ and $\{M_2(t_k), k=0,1,2,\ldots\}$ with rates $\delta_e$ and $\delta_r$, respectively. It follows that the SEIR Markov chain model $\{X(t_k), k=0,1,2,\ldots\}$ has the following transition probabilities:
 \begin{align*}
  &P(X(t_{k+1})=x^{k+1}|X(t_k)=x^k)\nonumber\\
  &= P((S(t_{k+1}),E(t_{k+1}),I(t_{k+1})=(s_{k+1},e_{k+1},i_{k+1})|X(t_k)=x^k)\nonumber\\
  &= \begin{cases}
  \binom{s_k}{s_{k+1}}(p^i(t_k))^{s_k-s_{k+1}}(1-p^i(t_k))^{s_{k+1}}\times\\
  \times \binom{e_k}{s_{k+1}+e_{k+1}}(P^i_{EI}(t_k))^{s_k+e_k-(s_{k+1}+e_{k+1})}\times\\ \times (1-P^i_{EI}(t_k))^{s_{k+1}+e_{k+1}-s_k}\times\\
  \times \binom{i_k}{s_{k+1}+e_{k+1}+i_{k+1}}(P^i_{IR}(t_k))^{s_k+e_k+i_k-(s_{k+1}+e_{k+1}+i_{k+1}}\times\\ \times (1-P^i_{IR}(t_k))^{s_{k+1}+e_{k+1}+i_{k+1}-(s_k+e_k)},&\text{for}\\
  &\text{$(s_{k+1},e_{k+1},i_{k+1})\in\Omega_X(t_{k+1})$}\\
  0, &\text{otherwise.}
  \end{cases}
  \end{align*}		
Moreover, the feasible region for the process is given as follows:
\begin{align*}
	\Omega_X(t_{k+1})=&\{(s_{k+1},e_{k+1},i_{k+1})\in\mathbb{Z}^3_{+}\vert s_k \le s_{k+1}+e_{k+1}\le s_k+e_k,\nonumber\\
	 &s_k+e_k \le s_{k+1}+e_{k+1}+i_{k+1}\le s_k+e_k+i_k\}.
	\end{align*}

\end{theorem}
\begin{proof}
The proof of this result is similar to Theorem~\ref{chp2.thm1}, and further analysis and details will appear in \cite{wanduku-diffusion}.
\end{proof}	
\subsection{Validation of the SEIR Markov chain models }
To validate the SEIR Markov chain epidemic models  $\{X(t_k):k\geq 0\}$ in  Theorem~\ref{chp2.thm1} and Theorem~\ref{chp.2.sec.2.1.7.inf.prob.thm1}, we provide some numerically simulated sample paths for the process $\{X(t_k):k\geq 0\}$, for selected values of $p$ and $\lambda$ to determine whether the process $\{X(t_k):k\geq 0\}$ represents a vital disease dynamics.

\begin{figure}[H]
	\centering
	\includegraphics[width=8cm]{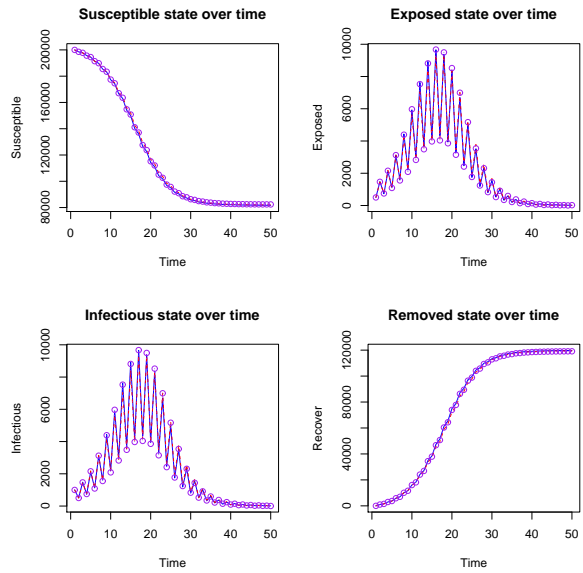}
	\caption{Shows three sample paths each for the states $S, E, I, R$ of the SEIR Markov chain model $\{X(t_k):k\geq 0\}$ with transition probabilities in Theorem~\ref{chp2.thm1}, whenever $p=0.15$,  and $\lambda=10$. In addition, the following initial conditions are used $S(t_{0})=200,000$, $E(t_{0})=500$, $I(t_{0})=1000$, and $R(t_{0})=0$. }
	\label{ch3.figure1}
\end{figure}

\begin{figure}[H]
	\centering
	\includegraphics[width=8cm]{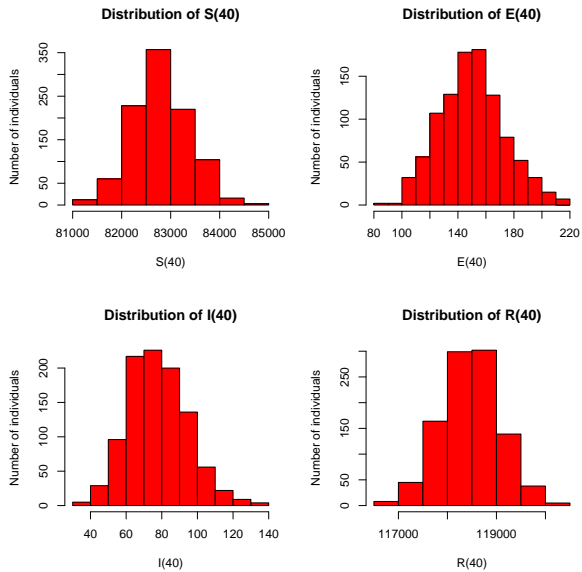}
	\caption{Shows the approximate distributions for the states $S, E, I, R$ of the SEIR Markov chain model $\{X(t_k):k\geq 0\}$ with transition probabilities in Theorem~\ref{chp2.thm1}, whenever $p=0.15$,  and $\lambda=10$. In addition, the following initial conditions are used $S(t_{0})=200,000$, $E(t_{0})=500$, $I(t_{0})=1000$, and $R(t_{0})=0$. The histograms are based on $1000$ sample realizations of the states $S, E, I, R$ at time $t_{40}$. Furthermore, the 95\% confidence intervals for the populations means of the states $S, E, I, R$ at time $t_{40}$ are respectively, $82766.66<E(S(t_{40}))<82838.89$, $149.1534 <E(E(t_{40}))<152.0026$, $77.25725<E(I(t_{40}))< 79.30675$ and $118431.7 <E(R(t_{40}))<118505$. }
	\label{ch3.figure2}
\end{figure}

Figure~\ref{ch3.figure1} depicts three sample paths each for the susceptible, exposed, infectious and removed states for the SEIR Markov chain epidemic model  $\{X(t_k):k\geq 0\}$ in  Theorem~\ref{chp2.thm1}. The following conditions are utilized:  $p=0.15$,  and $\lambda=10$. That is, from (\ref{p^i(t_k)}), the infectivity in the population is relatively low over time. In addition, the following initial conditions are used $S(t_{0})=200,000$, $E(t_{0})=500$, $I(t_{0})=1000$, and $R(t_{0})=0$. Moreover, histograms for the states $S, E, I, R$ of the process $\{X(t_k):k\geq 0\}$ based on 1000 sample realizations at the time $t_{40}$ are depicted in Figure~\ref{ch3.figure2}. Furthermore,  the 95\% confidence intervals for the populations means of the states $S, E, I, R$ at time $t_{40}$ are respectively, $82766.66<E(S(t_{40}))<82838.89$, $149.1534 <E(E(t_{40}))<152.0026$, $77.25725<E(I(t_{40}))< 79.30675$ and $118431.7 <E(R(t_{40}))<118505$.

Observe from Figure~\ref{ch3.figure1} that infectivity rises initially as more susceptible people become exposed, reaches a peak and decreases over time. The rise of the exposed state corresponds to a decrease of the susceptible state. The initial rise in the exposed state can be attributed to the initial high infectious population $I(t_{0})=1000$, and high initial state $E(t_{0})=500$ converting to the infectious state over the next time step. The exposed state reaches a peak and then decreases over time. Note that the pattern in the exposed state is translated to the infectious state, over the unit incubation period. And since the infectious state decreases over time and approaching zero, the infectivity in the population also decreases, and results  to lesser and lesser number of susceptible people infected. Infectivity slows down over time as nearly all infectious people receive treatment and recover from infection. Figure~\ref{ch3.figure2} shows that despite the fact that infectivity slows down, there are still significant amount of people in the exposed and infectious states on the $40^{th}$ epoch. Indeed, the 95\% confidence intervals for the states  $S, E, I, R$ at time $t_{40}$ are significantly large. This implies that infectivity continuous in the population over time, but at a lower rate.

\begin{figure}[H]
	\centering
	\includegraphics[width=8cm]{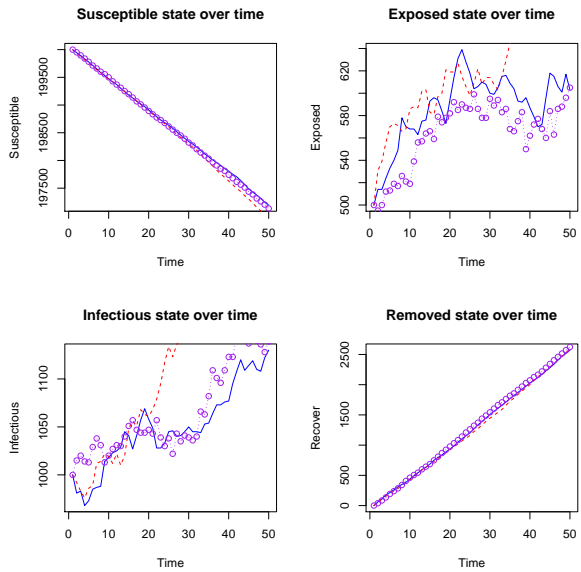}
	\caption{Shows three sample paths each for the states $S, E, I, R$ of the SEIR Markov chain model $\{X(t_k):k\geq 0\}$ with transition probabilities in Theorem~\ref{chp.2.sec.2.1.7.inf.prob.thm1}, whenever $p=0.0055$,   $\lambda=10$, the average incubation $T_{1}$ and infectious $T_{2}$ periods are respectively, $E(T_{1})=10$ and $E(T_{2})=20$. In addition, the following initial conditions are used $S(t_{0})=200,000$, $E(t_{0})=500$, $I(t_{0})=1000$, and $R(t_{0})=0$. }
	\label{ch3.figure3}
\end{figure}

\begin{figure}[H]
	\centering
	\includegraphics[width=8cm]{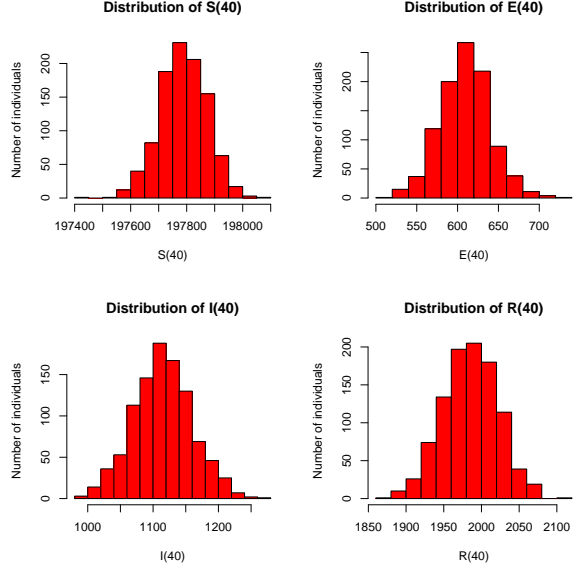}
	\caption{Shows the approximate distributions for the states $S, E, I, R$ of the SEIR Markov chain model $\{X(t_k):k\geq 0\}$ with transition probabilities in Theorem~\ref{chp.2.sec.2.1.7.inf.prob.thm1}, whenever $p=0.0055$,   $\lambda=10$, the average incubation $T_{1}$ and infectious $T_{2}$ periods are respectively, $E(T_{1})=10$ and $E(T_{2})=20$. In addition, the following initial conditions are used $S(t_{0})=200,000$, $E(t_{0})=500$, $I(t_{0})=1000$, and $R(t_{0})=0$. The histograms are based on $1000$ sample realizations of the states $S, E, I, R$ at time $t_{40}$. Furthermore, the 95\% confidence intervals for the populations means of the states $S, E, I, R$ at time $t_{40}$ are respectively, $197783.9<E(S(t_{40}))<197794.2$, $607.8549 <E(E(t_{40}))<611.7331$, $1112.502<E(I(t_{40}))< 1118.106$ and $1983.614 <E(R(t_{40}))<1988.112$. }
	\label{ch3.figure4}
\end{figure}

Figure~\ref{ch3.figure3} depicts three sample paths each for the susceptible, exposed, infectious and removed states for the SEIR Markov chain epidemic model  $\{X(t_k):k\geq 0\}$ in  Theorem~\ref{chp.2.sec.2.1.7.inf.prob.thm1}. The following conditions are utilized:  $p=0.0055$,   $\lambda=10$, the average incubation $T_{1}$ and infectious $T_{2}$ periods are respectively, $E(T_{1})=10$ and $E(T_{2})=20$. That is, from (\ref{p^i(t_k)}), the infectivity in the population is relatively rising over time. In addition, the following initial conditions are used $S(t_{0})=200,000$, $E(t_{0})=500$, $I(t_{0})=1000$, and $R(t_{0})=0$. Moreover, histograms for the states $S, E, I, R$ of the process $\{X(t_k):k\geq 0\}$ based on 1000 sample realizations at the time $t_{40}$ are depicted in Figure~\ref{ch3.figure4}. Furthermore,  the 95\% confidence intervals for the populations means of the states $S, E, I, R$ at time $t_{40}$ are respectively, $197783.9<E(S(t_{40}))<197794.2$, $607.8549 <E(E(t_{40}))<611.7331$, $1112.502<E(I(t_{40}))< 1118.106$ and $1983.614 <E(R(t_{40}))<1988.112$.

Observe from Figure~\ref{ch3.figure3} that infectivity generally rises over time as more susceptible people become exposed over time. The rise in the exposed state corresponds to a continuous decrease in the susceptible state, and also corresponds to a continuous rise in the infectious state as more exposed people develop full-blown disease and become infectious. Note that the average infectious period $E(T_{2})=20$ is twice the incubation period  $E(T_{1})=10$, implying that more people tend to develop-full blown disease, than they recover from disease. Furthermore, since the incubation is no longer fixed as in Figure~\ref{ch3.figure1}, there is no translation from the exposed class to the infectious state. The recovery from disease occurs at a nearly steady rate, and rises over time as more infectious people become removed. Figure~\ref{ch3.figure4} shows that with the rising infectivity in the population, there are still significant amounts of people in the susceptible state who have never been infected on the $40^{th}$ epoch. Indeed, the 95\% confidence intervals for the states  $S, E, I, R$ at time $t_{40}$ are significantly large. These intervals suggest that infectivity continuous in the population over time, and at a higher rate, since there are still significant number of people in the exposed and infectious states at time $t_{40}$.


\section{PARAMETER ESTIMATION}\label{chp4}
In this section, we find estimators for the true parameters of our SEIR Markov chain model using observed data for the state of the process over time. Utilizing similar ideas  in \cite{ fierro-victoria,yae, ross-tb,keeling-1}, we find maximum likelihood estimators\cite{cb} for the probability of passing infection to a susceptible person after one interaction with an infectious person at any time $t_k$, $p$, and the average number of people a susceptible  individual meets and interacts with per unit time, $\lambda$, for the SEIR Markov chain model $\{X(t_k):k\geq 0\}$ in the case where the transition probabilities are defined in Theorem~\ref{chp2.thm1}.

 Indeed, note that the parameter $\Theta = (p,\lambda)$ represents fixed measures in the population at each time $t_k$, that is, $p$ and $\lambda$ represent fixed measurements for events occurring in the population during every epoch $k$ $([t_k,t_{k+1}))$, where the population at any time $t_k, k\geq 0s$ is defined by the random vector
\begin{align}\label{chp3.sec0.X(t_k)}
X(t_k)= (S(t_k), E(t_k), I(t_k)).
\end{align}
Let $\hat{x}(t_k)$ be the observed value of the random vector $X(t_k)$ at any time $t_k, k=0,1,2,\ldots$ defined in (\ref{chp3.sec0.X(t_k)}). That is,
\begin{align}\label{chp3.sec0.hatx}
\hat{x}(t_k)=(\hat{s}_k, \hat{e}_k, \hat{i}_k), \forall k=0,1,2,\ldots,
\end{align}
where $\hat{s}_k, \hat{e}_k, \hat{i}_k \in \mathbb{Z}_{+} $ are non-negative observed constant values for each component of $X(t_k)$, at any time $t_k, k=0,1,2,\ldots$.\par
\begin{figure}[H]
	\centering
	\includegraphics[width=10cm]{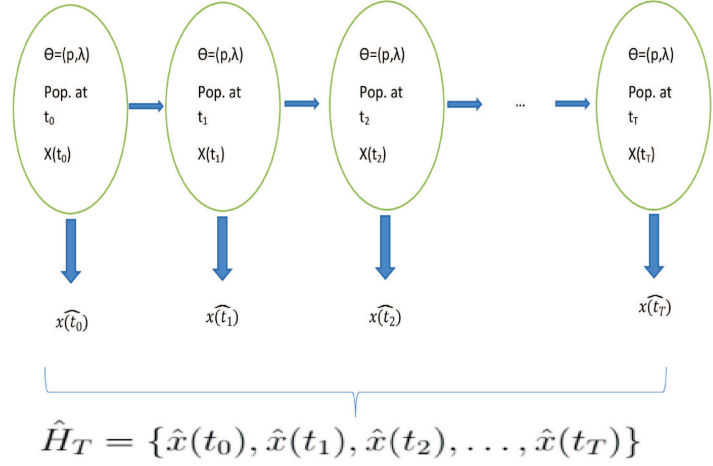}
	\caption{Shows the transition of the process $\{X(t_k);k=0,1,2,\ldots\}$ over time $k=0,1,2,\ldots,T$, and observed data $\hat{H}_T= \{\hat{x}(t_0), \hat{x}(t_1), \hat{x}(t_2),\ldots,\hat{x}(t_T)\}$. The parameters $\Theta =(p,\lambda)$ are constant in the population at all times $k=0,1,2,\ldots,T$.}
	\label{Fig 2}
\end{figure}

The population $X(t_k)$ is observed over the time units, $t_k, k=0,1,2,\ldots,T$, where the initial state $X(t_0)= \hat{x}(t_0)$ is assumed to be known. That is, $X(t_0)$ is deterministic, and the observed data consists of the measurements
\begin{align}
\hat{x}(t_0), \hat{x}(t_1), \hat{x}(t_2),\ldots,\hat{x}(t_T).
\end{align}
We define the finite collection of random variables $X(t_0),X(t_1),X(t_2),\ldots,X(t_T)$ representing the population over times $t_k, k=0,1,2,\ldots,T$ as follows:
\begin{align}\label{chp3.sec0.HT}
H_T=\{X(t_0),X(t_1),X(t_2),\ldots,X(t_T)\},
\end{align}
and from (\ref{chp3.sec0.hatx}), the observed values of $H_T$ are given as,
\begin{align}\label{chp3.sec0.hatht}
\hat{H}_T= \{\hat{x}(t_0), \hat{x}(t_1), \hat{x}(t_2),\ldots,\hat{x}(t_T)\}.
\end{align}
We use the observed sample path $\hat{H}_T$ of the process $\{X(t_k);k=0,1,2,\ldots\}$ to find the maximum likelihood estimates for the parameters $\Theta =(p,\lambda)$. The generation of the sample path $\hat{H}_T$ from the population $X(t_k)$ over the times $k=0,1,2,\ldots,T$ is illustrated in Figure \ref{Fig 2}.
%

We assume that we have data for the SEIR infectious disease such as Pneumonia or influenza over time units $t_k, k=0,1,2,\ldots,T$ denoted $\hat{H}_T$, where $\hat{H}_T$ is defined in (\ref{chp3.sec0.hatht}), and $\hat{H}_T$ is one realization of the human population over time denoted $H_T$, defined in (\ref{chp3.sec0.HT}). Furthermore, we assume the SEIR model has transition probabilities in Theorem~\ref{chp2.thm1}. From (\ref{chp3.sec0.hatx}), (\ref{chp3.sec0.HT}), and (\ref{chp3.sec0.hatht}), the likelihood function of $\Theta=(p,\lambda)$ is defined as follows:
\begin{align}\label{chp3.sec.3.1.L}
L(\Theta|\hat{H}_T)&= L(p,\lambda|\hat{H}_T)= P(H_T=\hat{H}_T|p,\lambda)\nonumber\\
&= P(X(t_T)=\hat{x}(t_T),X(t_{T-1})=\hat{x}(t_{T-1}),\ldots,X(t_0)=\hat{x}(t_0)|p,\lambda).
\end{align}
From (\ref{chp3.sec.3.1.L}), applying the multiplication rule, it is easy to see that
\begin{align}\label{chp3.sec.3.1.LM}
L(p,\lambda|\hat{H}_T)&=P(X(t_T)=\hat{x}(t_T)|X(t_{T-1})=\hat{x}(t_{T-1}),\ldots,X(t_0)=\hat{x}(t_0);p,\lambda)\times\nonumber\\
\times &P(X(t_{T-1})=\hat{x}(t_{T-1})|X(t_{T-2})=\hat{x}(t_{T-2}),\ldots,X(t_0)=\hat{x}(t_0);p,\lambda)\times \nonumber\\
&\vdots\nonumber\\
\times &P(X(t_1)=\hat{x}(t_1)|X(t_0)=\hat{x}(t_0);p,\lambda)\times P(X(t_0)=\hat{x}(t_0);p,\lambda).
\end{align}
But, since $\{X(t_k), k=0,1,2,\ldots\}$ is a Markov chain, and since it is assumed $X(t_0)$ is known, it is easy to see that (\ref{chp3.sec.3.1.LM}) reduces to
\begin{align}\label{chp3.sec3.1.L1}
L(p,\lambda|\hat{H}_T)= \prod_{k=1}^T P(X(t_k)=\hat{x}(t_k)|X(t_{k-1})=\hat{x}(t_{k-1});p,\lambda).
\end{align}
It follows from (\ref{chp3.sec3.1.L1}), Theorem \ref{chp2.thm1},
\begin{align}\label{chp3.sec3.1.L2}
L(p,\lambda|\hat{H}_T)= \prod_{k=1}^T P(S(t_k)=\hat{s}_k,E(t_k)=\hat{e}_k,I(t_k)=\hat{i}_k|X(t_{k-1})=\hat{x}(t_{k-1});p,\lambda).
\end{align}
The equation (\ref{chp3.sec3.1.L2}) is the likelihood function with respect to the parameters $p$ and $\lambda$.
We note that applying the maximization technique to find the MLE's $\hat{p}$, and $\hat{\lambda}$, for $p$, and $\lambda$, respectively, using the likelihood function $L$ defined in (\ref{chp3.sec3.1.L2}) leads to  intractable equations for the derivatives of the log-likelihood of $L$ with respect to $p$, or $\lambda$, set to zero. Thus, we apply the expectation maximization (EM) algorithm to find an appropriate MLE for $p$- the probability of passing infection to a susceptible person after one intersection with an infection person, and for $\lambda$- the average number of people an individual meets per unit time.
\subsection{The EM Algorithm and Jensen's Inequality}
We recall the following. The Expectation Maximization (EM) algorithm is an iterative algorithm used to find the MLE of a parameter $\Theta$ of a given distribution \cite{mgp, jba}. There are two cases where the algorithm is most useful: (1) when the data available for maximum likelihood estimation technique has missing components, and (2) when maximizing the likelihood function leads to an intractable equation, but adding missing data can simplify the process. It is for the second case in our problem that we utilize the EM algorithm.

Suppose we have observed data $Y$, and likelihood function $L(\Theta|Y)=P(Y|\Theta)$, and suppose the vector $Z$ is missing data or a missing component, so that $X= (Y,Z)$ is the complete data. The complete log-likelihood function $log (L(\Theta|X))=log (P(Y,Z|\Theta))$ is obtained and maximized to find the MLE of $\Theta$ in two basic algorithm steps, namely- the expectation (E)-step, and the maximization (M)-step.\par

The E-step consists of finding the expected value of the complete log-likelihood function 
\begin{align}\label{chp3.sec.3.1.EZ}
E_{Z|Y;\Theta}[log (L(\Theta|X))]&= E_{Z|Y;\Theta}[log (P(Y,Z|\Theta))]\nonumber\\
&=\sum_{Z}log (P(Y,Z|\Theta) P(Z|Y;\Theta)).
\end{align}
The M-step consists of maximizing $E_{Z|Y;\Theta}[log (L(\Theta|X))]$ to find an estimate $\hat{\Theta}$ for $\Theta$.
This process is summarized in the following steps:
\begin{enumerate}
\item Let $m=0$ and $\hat{\Theta}^m$ be an initial guess for $\Theta$.\label{chp3.item1}
\item Given the observed data $Y,$ and assuming that the guess $\hat{\Theta}^m$ is correct, calculate the conditional probability distribution $P(Z|Y,\hat{\Theta}^m)$ for the missing data $Z$.
\item Find the conditional expected log-likelihood referred to as $Q$, that is,
\begin{align}
Q(\Theta|\hat{\Theta}^m)&=\sum_{Z}log(P(Y,Z|\Theta)P(Z|Y,\hat{\Theta}^m))\nonumber\\
&=E_{Z|Y,\hat{\Theta}^m}[log(P(X|\Theta))],
\end{align}
where $X=(Y,Z)$.
\item Find the $\Theta$ that maximizes $Q(\Theta|\hat{\Theta}^m)$. The result will be the new $\hat{\Theta}^{m+1}$. That is ,\label{chp3.item2}
\begin{align}
\hat{\Theta}^{m+1}= argmax_\Theta Q(\Theta|\hat{\Theta}^m)
\end{align}
\item Update $\hat{\Theta}^m$ and repeat step \ref{chp3.item1} to step \ref{chp3.item2} until $\Theta$ stops noticeably changing.
\end{enumerate}
The E-step can be obtained by applying Jensen's inequality. We recall Jensen's inequality \cite{cb} in the following:
\begin{lemma}
Suppose f is a convex function, and X is a random variable, then
\begin{align}
E[f(X)]\ge f(E[X]).
\end{align}
Conversely, if you have a concave function (e.g. a logarithmic function), then
\begin{align}
E[f(X)]\le f(E[X]).
\end{align}
\end{lemma}
From (\ref{chp3.sec.3.1.EZ}), let $Y=\hat{H}_T$ represent the observed data defined in (\ref{chp3.sec0.hatht}). The following random missing information Z are incorporated to make the log-likelihood function $log(L)$ more tractable, where L is given in (\ref{chp3.sec3.1.L2}). We utilize  Assumption~\ref{chp2.sec2.assum2}.
\begin{enumerate}
\item Suppose the $i^{th}$ susceptible person meets $f^i_{t_k} = N$ discrete random number of people during the epoch $k$ $([t_k,t_{k+1}))$ ($i.e.$ over epoch: $k=0,1,\ldots,T$) with rate $\lambda $. Define the collection $\vec{f}^i_T = \{f^i_{t_0},f^i_{t_1},\ldots,f^i_{t_k},\ldots,f^i_{t_T}\}$, where $k\in \{0,1,2,\ldots,T\}$ and $i\in \{1,2,\ldots,s_k\}$. From Assumption~\ref{chp2.sec2.assum2}, for each $k= 0,1,2,\ldots,$ and $i\in \{1,2,\ldots,s_k\}$,
     \begin{align}\label{chp3.sec3.1.fiN}
f^i_{t_k}= N, \forall N\ge 0,\quad and \quad N \sim Poisson (\lambda \Delta t).
\end{align}

\item The collection $\vec{y}^i_{TN} = \{y^i_{t_0N},y^i_{t_1N},\ldots,y^i_{t_kN},\dots,y^i_{t_TN}\}$, for each $k\in \{0,1,2,\ldots,T\}$ and $i\in \{1,2,\ldots,s_k\}$. Given that the $i^{th}$ susceptible person meets $f^i_{t_k} = N$ people during epoch $k$, let $y^i_{t_kN}$ be a discrete random variable representing the number of infectious people among the $N$ people. Therefore,
\begin{align}\label{chp3.sec3.1.yiN}
y^i_{t_kN}=j,j=0,1,2,\ldots,N.
\end{align}
\item Given $f^i_{t_k} = N$ people met at time $t_k$, and also given $y^i_{t_kN}=j$ infectious people present among the $N$ people, let
\begin{align}\label{chp3.sec3.1.ditk}
d^i_{t_kNj}=l,l=1,2,\ldots,j,
\end{align}
be a categorical random variable (indicator random variable) representing the $l^{th}$ infectious person who passes infection at time $t_k$. Then the collection $\vec{d}^i_{TNj}$ can be represent as $\vec{d}^i_{TNj}=\{d^i_{t_0Nj},d^i_{t_1Nj},\ldots,d^i_{t_kNj},\ldots,d^i_{t_TNj}\}$, where $N=0,1,2,\ldots$ and $j=0,1,2,\ldots,N$.
\end{enumerate}
We consider a step-by-step approach to add the random missing data $\vec{f}^i_T$, $\vec{y}^i_{TN}$ and $\vec{d}^i_{TNj}$ into the incomplete likelihood function $L$, defined in (\ref{chp3.sec3.1.L2}).
\begin{lemma}\label{chp3.sec3.1.1.lema8}
Given the missing information $\vec{f}^i_T, \vec{y}^i_{TN}$ and $\vec{d}^i_{TNj}$ defined in (\ref{chp3.sec3.1.fiN}), (\ref{chp3.sec3.1.yiN}) and (\ref{chp3.sec3.1.ditk}), then the log-likelihood function $log L(p,\lambda|\hat{H}_T)$ satisfies the following inequality:
\begin{align}
&log L(p,\lambda|\hat{H}_T)\ge \sum_{k=1}^{T}\sum_{N=0}^{\infty}\sum_{j=0}^{N}\sum_{l=1}^{j}\big[P(f^i_{t_k}=N|X(t_{k-1})=\hat{x}(t_{k-1});p,\lambda)\times \nonumber\\
&\times P(y^i_{t_kN}=j|f^i_{t_k}=N,X(t_{k-1})=\hat{x}(t_{k-1});p,\lambda)\times\nonumber\\
&\times P(d^i_{t_kNj}=l|f^i_{t_k}=N,y^i_{t_kN}=j,X(t_{k-1})=\hat{x}(t_{k-1});p,\lambda)\times\nonumber\\
&\times log\{P(S(t_k)=\hat{s}_k,f^i_{t_k}=N,y^i_{t_kN}=j,d^i_{t_kNj}=l|X(t_{k-1})=\hat{x}(t_{k-1});p,\lambda)\}\big]\nonumber\\
&+ \sum_{k=1}^{T}\sum_{N=0}^{\infty}\sum_{j=0}^{N}\sum_{l=1}^{j}\big[P(f^i_{t_k}=N|S(t_k)=\hat{s}_k,X(t_{k-1})=\hat{x}(t_{k-1});p,\lambda)\times\nonumber\\
&\times P(y^i_{t_kN}=j|f^i_{t_k}=N,S(t_k)=\hat{s}_k,X(t_{k-1})=\hat{x}(t_{k-1});p,\lambda)\times \nonumber\\
&\times P(d^i_{t_kNj}=l|f^i_{t_k}=N,y^i_{t_kN}=j,S(t_k)=\hat{s}_k,X(t_{k-1})=\hat{x}(t_{k-1});p,\lambda)\times \nonumber\\
&\times log\{P(E(t_k)=\hat{e}_k,f^i_{t_k}=N,y^i_{t_kN}=j,d^i_{t_kNj}=l|S(t_k)=\hat{s}_k,X(t_{k-1})=\hat{x}(t_{k-1});p,\lambda)\}\big] \nonumber\\
&+ \sum_{k=1}^{T}\sum_{N=0}^{\infty}\sum_{j=0}^{N}\sum_{l=1}^{j}\big[P(f^i_{t_k}=N|S(t_k)=\hat{s}_k,E(t_k)=\hat{e}_k,X(t_{k-1})=\hat{x}(t_{k-1});p,\lambda)\times\nonumber\\
&\times P(y^i_{t_kN}=j|f^i_{t_k}=N,S(t_k)=\hat{s}_k,E(t_k)=\hat{e}_k,X(t_{k-1})=\hat{x}(t_{k-1});p,\lambda)\times\nonumber\\
&\times P(d^i_{t_kNj}=l|f^i_{t_k}=N,y^i_{t_kN}=j,S(t_k)=\hat{s}_k,E(t_k)=\hat{e}_k,X(t_{k-1})=\hat{x}(t_{k-1});p,\lambda)\times \nonumber\\
&\times log\{P(I(t_k)=\hat{i}_k,f^i_{t_k}=N,y^i_{t_kN}=j,d^i_{t_kNj}=l|S(t_k)=\hat{s}_k,E(t_k)=\hat{e}_k,\nonumber\\
&X(t_{k-1})=\hat{x}(t_{k-1});p,\lambda)\}\big]- \chi_1-\chi_2-\chi_3,\label{chp3.sec3.1.L0}
\end{align}
where $\chi_1, \chi_2$, and $\chi_3$ are probability terms that depend on $f^i_{t_k}$, $y^i_{t_kN}$ and $d^i_{t_kNj}$.
\end{lemma}
\begin{proof}
From (\ref{chp3.sec3.1.L2}), denote the log-likelihood $l(p,\lambda|\hat{H}_T)\equiv log (L(p,\lambda|\hat{H}_T))$. It follows from (\ref{chp3.sec3.1.L2}) that adding the missing random data $\vec{f}^i_T$, we obtain
\begin{align}
l(p,\lambda|\hat{H}_T)&=log \prod_{k=1}^T P(S(t_k)=\hat{s}_k,E(t_k)=\hat{e}_k,I(t_k)=\hat{i}_k|X(t_{k-1})=\hat{x}(t_{k-1});p,\lambda)\nonumber\\
&=\sum_{k=1}^{T}log \bigg[\sum_{N=0}^{\infty}P(S(t_k)=\hat{s}_k,f^i_{t_k}=N|X(t_{k-1})=\hat{x}(t_{k-1});p,\lambda)\bigg]\nonumber\\
&+ \sum_{k=1}^{T}log \bigg[\sum_{N=0}^{\infty}P(E(t_k)=\hat{e}_k,f^i_{t_k}=N|S(t_k)=\hat{s}_k,X(t_{k-1})=\hat{x}(t_{k-1});p,\lambda)\bigg]\nonumber\\
&+ \sum_{k=1}^{T}log \bigg[\sum_{N=0}^{\infty}P(I(t_k)=\hat{i}_k,f^i_{t_k}=N|S(t_k)=\hat{s}_k,E(t_k)=\hat{e}_k,\nonumber\\
&X(t_{k-1})=\hat{x}(t_{k-1});p,\lambda)\bigg].\label{chp3.sec3.1.L3}
\end{align}
Applying algebraic manipulations and Jensen's inequality to the three summation components of  (\ref{chp3.sec3.1.L3}), leads to the following,
\begin{align}
&l(p,\lambda|\hat{H}_T)\ge \sum_{k=1}^{T}\sum_{N=0}^{\infty}log\{P(S(t_k)=\hat{s}_k,f^i_{t_k}=N|X(t_{k-1})=\hat{x}(t_{k-1});p,\lambda)\}\times \nonumber\\
&\times P(f^i_{t_k}=N|X(t_{k-1})=\hat{x}(t_{k-1});p,\lambda)\nonumber\\
&+ \sum_{k=1}^{T}\sum_{N=0}^{\infty}log\{P(E(t_k)=\hat{e}_k,f^i_{t_k}=N|S(t_k)=\hat{s}_k,X(t_{k-1})=\hat{x}(t_{k-1});p,\lambda)\}\times \nonumber\\
&\times P(f^i_{t_k}=N|S(t_k)=\hat{s}_k,X(t_{k-1})=\hat{x}(t_{k-1});p,\lambda)\nonumber\\
&+ \sum_{k=1}^{T}\sum_{N=0}^{\infty}log\{P(I(t_k)=\hat{i}_k,f^i_{t_k}=N|S(t_k)=\hat{s}_k,E(t_k)=\hat{e}_k,X(t_{k-1})=\hat{x}(t_{k-1});p,\lambda)\}\nonumber\\
&\times P(f^i_{t_k}=N|S(t_k)=\hat{s}_k,E(t_k)=\hat{e}_k,X(t_{k-1})=\hat{x}(t_{k-1});p,\lambda)- \chi_1,\label{chp3.sec3.1.L5}
\end{align}
where $\chi_1$ in (\ref{chp3.sec3.1.L5}) is given as follows,
\begin{align}
&\chi_1= \sum_{k=1}^{T}\sum_{N=0}^{\infty}log \{P(f^i_{t_k}=N|X(t_{k-1})=\hat{x}(t_{k-1});p,\lambda)\}\times P(f^i_{t_k}=N|\nonumber\\
&X(t_{k-1})=\hat{x}(t_{k-1});p,\lambda)+ \sum_{k=1}^{T}\sum_{N=0}^{\infty}log\{P(f^i_{t_k}=N|S(t_k)=\hat{s}_k,X(t_{k-1})=\hat{x}(t_{k-1});p,\lambda)\} \nonumber\\
&\times P(f^i_{t_k}=N|S(t_k)=\hat{s}_k,X(t_{k-1})=\hat{x}(t_{k-1});p,\lambda)\nonumber\\
& + \sum_{k=1}^{T}\sum_{N=0}^{\infty}log \{P(f^i_{t_k}=N|S(t_k)=\hat{s}_k,E(t_k)=\hat{e}_k,X(t_{k-1})=\hat{x}(t_{k-1});p,\lambda)\}\times\nonumber\\
&\times P(f^i_{t_k}=N|S(t_k)=\hat{s}_k,E(t_k)=\hat{e}_k,X(t_{k-1})=\hat{x}(t_{k-1});p,\lambda).\label{chp3.sec3.1.L6}
\end{align}
We add missing data $\vec{y}^i_{TN}$ in (\ref{chp3.sec3.1.yiN}) into the partially complete log-likelihood function $log\{P(S(t_k)=\hat{s}_k,f^i_{t_k}=N|X(t_{k-1})=\hat{x}(t_{k-1});p,\lambda)\}$, \\
$log\{P(E(t_k)=\hat{e}_k,f^i_{t_k}=N|S(t_k)=\hat{s}_k,X(t_{k-1})=\hat{x}(t_{k-1});p,\lambda)\}$ and \\
$log\{P(I(t_k)=\hat{i}_k,f^i_{t_k}=N|S(t_k)=\hat{s}_k,E(t_k)=\hat{e}_k$\}, $\forall k\in \{1,2,\ldots,T\}$;\\ $N\in \{0,1,2,\ldots,N\}$, and apply the same technique in (\ref{chp3.sec3.1.L3})-(\ref{chp3.sec3.1.L6}), as follows.\par
From (\ref{chp3.sec3.1.L5}), it is easy to see that
\begin{align}
&l(p,\lambda|\hat{H}_T)\ge \sum_{k=1}^{T}\sum_{N=0}^{\infty}log\{\sum_{j=0}^{N}P(S(t_k)=\hat{s}_k,f^i_{t_k}=N, y^i_{t_kN}=j|X(t_{k-1})=\hat{x}(t_{k-1});\nonumber\\
&p,\lambda)\} \times P(f^i_{t_k}=N|X(t_{k-1})=\hat{x}(t_{k-1});p,\lambda)\nonumber\\
&+ \sum_{k=1}^{T}\sum_{N=0}^{\infty}log\{\sum_{j=0}^{N}P(E(t_k)=\hat{e}_k,f^i_{t_k}=N,y^i_{t_kN}=j|S(t_k)=\hat{s}_k,X(t_{k-1})=\hat{x}(t_{k-1});\nonumber\\
&p,\lambda)\}\times P(f^i_{t_k}=N|S(t_k)=\hat{s}_k,X(t_{k-1})=\hat{x}(t_{k-1});p,\lambda)\nonumber\\
&+ \sum_{k=1}^{T}\sum_{N=0}^{\infty}log\{\sum_{j=0}^{N}P(I(t_k)=\hat{i}_k,f^i_{t_k}=N,y^i_{t_kN}=j|S(t_k)=\hat{s}_k,E(t_k)=\hat{e}_k,\nonumber\\
&X(t_{k-1})=\hat{x}(t_{k-1});p,\lambda)\}\times P(f^i_{t_k}=N|S(t_k)=\hat{s}_k,E(t_k)=\hat{e}_k,X(t_{k-1})=\hat{x}(t_{k-1});\nonumber\\
&p,\lambda)- \chi_1.\label{chp3.sec3.1.L7}
\end{align}
Applying similar algebraic manipulations and Jensen's inequality to (\ref{chp3.sec3.1.L7}), leads to the following,
\begin{align}
&l(p,\lambda|\hat{H}_T)\ge \sum_{k=1}^{T}\sum_{N=0}^{\infty}\sum_{j=0}^{N}log\{P(S(t_k)=\hat{s}_k,f^i_{t_k}=N,y^i_{t_kN}=j|X(t_{k-1})=\hat{x}(t_{k-1});\nonumber\\
&p,\lambda)\}\times P(y^i_{t_kN}=j|f^i_{t_k}=N,X(t_{k-1})=\hat{x}(t_{k-1});p,\lambda)\times P(f^i_{t_k}=N|\nonumber\\
&X(t_{k-1})=\hat{x}(t_{k-1});p,\lambda)+ \sum_{k=1}^{T}\sum_{N=0}^{\infty}\sum_{j=0}^{N}log\{P(E(t_k)=\hat{e}_k,f^i_{t_k}=N,y^i_{t_kN}=j|S(t_k)=\hat{s}_k,\nonumber\\
&X(t_{k-1})=\hat{x}(t_{k-1});p,\lambda)\}\times  P(y^i_{t_kN}=j|f^i_{t_k}=N,S(t_k)=\hat{s}_k,\nonumber\\&
X(t_{k-1})=\hat{x}(t_{k-1});p,\lambda)\times
P(f^i_{t_k}=N|S(t_k)=\hat{s}_k,X(t_{k-1})=\hat{x}(t_{k-1});p,\lambda)\nonumber\\&+ \sum_{k=1}^{T}\sum_{N=0}^{\infty}\sum_{j=0}^{N}log\{P(I(t_k)=\hat{i}_k,f^i_{t_k}=N,y^i_{t_kN}=j|S(t_k)=\hat{s}_k,E(t_k)=\hat{e}_k,\nonumber\\&X(t_{k-1})=\hat{x}(t_{k-1});p,\lambda)\}\times P(y^i_{t_kN}=j|f^i_{t_k}=N,S(t_k)=\hat{s}_k,E(t_k)=\hat{e}_k,\nonumber\\
&X(t_{k-1})=\hat{x}(t_{k-1});p,\lambda)\times P(f^i_{t_k}=N|S(t_k)=\hat{s}_k,E(t_k)=\hat{e}_k,X(t_{k-1})=\hat{x}(t_{k-1});p,\lambda)\nonumber\\
&- \chi_1-\chi_2,\label{chp3.sec3.1.L8}
\end{align}
where $\chi_2$ in (\ref{chp3.sec3.1.L8}) is given as follows,
\begin{align}
&\chi_2= \sum_{k=1}^{T}\sum_{N=0}^{\infty}\sum_{j=0}^{N}log \{P(y^i_{t_kN}=j|f^i_{t_k}=N,X(t_{k-1})=\hat{x}(t_{k-1});p,\lambda)\}\times\nonumber\\
&\times P(y^i_{t_kN}=j|f^i_{t_k}=N,X(t_{k-1})=\hat{x}(t_{k-1});p,\lambda)\times P(f^i_{t_k}=N|X(t_{k-1})=\hat{x}(t_{k-1});p,\lambda)\nonumber\\
& + \sum_{k=1}^{T}\sum_{N=0}^{\infty}\sum_{j=0}^{N}log\{P(y^i_{t_kN}=j|f^i_{t_k}=N,S(t_k)=\hat{s}_k,X(t_{k-1})=\hat{x}(t_{k-1});p,\lambda)\}\times \nonumber\\
&\times P(y^i_{t_kN}=j|f^i_{t_k}=N,S(t_k)=\hat{s}_k,X(t_{k-1})=\hat{x}(t_{k-1});p,\lambda)\times\nonumber\\
&\times P(f^i_{t_k}=N|S(t_k)=\hat{s}_k,X(t_{k-1})=\hat{x}(t_{k-1});p,\lambda)\nonumber\\
& + \sum_{k=1}^{T}\sum_{N=0}^{\infty}\sum_{j=0}^{N}log \{P(y^i_{t_kN}=j|f^i_{t_k}=N,S(t_k)=\hat{s}_k,E(t_k)=\hat{e}_k,X(t_{k-1})=\hat{x}(t_{k-1});\nonumber\\
&p,\lambda)\}\times P(y^i_{t_kN}=j|f^i_{t_k}=N,S(t_k)=\hat{s}_k,E(t_k)=\hat{e}_k,X(t_{k-1})=\hat{x}(t_{k-1});p,\lambda)\times \nonumber\\
&\times P(f^i_{t_k}=N|S(t_k)=\hat{s}_k,E(t_k)=\hat{e}_k,X(t_{k-1})=\hat{x}(t_{k-1});p,\lambda).\label{chp3.sec3.1.L9}
\end{align}
Similarly, we add missing data $\vec{d}^i_{TNj}$ in (\ref{chp3.sec3.1.ditk}) into the partially complete log-likelihood function
$log\{P(S(t_k)=\hat{s}_k,f^i_{t_k}=N,y^i_{t_kN}=j|X(t_{k-1})=\hat{x}(t_{k-1});p,\lambda)\}$, \\
$log\{P(E(t_k)=\hat{e}_k,f^i_{t_k}=N,y^i_{t_kN}=j|S(t_k)=\hat{s}_k,X(t_{k-1})=\hat{x}(t_{k-1});p,\lambda)\}$ and \\
$log\{P(I(t_k)=\hat{i}_k,f^i_{t_k}=N,y^i_{t_kN}=j|S(t_k)=\hat{s}_k,E(t_k)=\hat{e}_k,X(t_{k-1})=\hat{x}(t_{k-1});p,\lambda)\}$, $\forall k\in \{1,2,\ldots,T\}$; $N\in \{0,1,2,\ldots,N\}$, and $j\in \{0,1,2,\ldots,N\}$, and apply the same technique in (\ref{chp3.sec3.1.L7})-(\ref{chp3.sec3.1.L9}), as follows.\par
From (\ref{chp3.sec3.1.L8}),
\begin{align}
&l(p,\lambda|\hat{T})\ge \sum_{k=1}^{T}\sum_{N=0}^{\infty}\sum_{j=0}^{N}log\{\sum_{l=1}^{j}P(S(t_k)=\hat{s}_k,f^i_{t_k}=N,y^i_{t_kN}=j,d^i_{t_kNj}=l|\nonumber\\
&X(t_{k-1})=\hat{x}(t_{k-1});p,\lambda)\}\times P(y^i_{t_kN}=j|f^i_{t_k}=N,X(t_{k-1})=\hat{x}(t_{k-1});p,\lambda)\times P(f^i_{t_k}=N|\nonumber\\
&X(t_{k-1})=\hat{x}(t_{k-1});p,\lambda)+ \sum_{k=1}^{T}\sum_{N=0}^{\infty}\sum_{j=0}^{N}log\{\sum_{l=1}^{j}P(E(t_k)=\hat{e}_k,f^i_{t_k}=N,y^i_{t_kN}=j,\nonumber\\
&d^i_{t_kNj}=l|S(t_k)=\hat{s}_k,X(t_{k-1})=\hat{x}(t_{k-1});p,\lambda)\}\times
P(y^i_{t_kN}=j|f^i_{t_k}=N,S(t_k)=\hat{s}_k,\nonumber\\
&X(t_{k-1})=\hat{x}(t_{k-1});p,\lambda)\times P(f^i_{t_k}=N|S(t_k)=\hat{s}_k,X(t_{k-1})=\hat{x}(t_{k-1});p,\lambda)\nonumber\\
&+ \sum_{k=1}^{T}\sum_{N=0}^{\infty}\sum_{j=0}^{N}log\{\sum_{l=1}^{j}P(I(t_k)=\hat{i}_k,f^i_{t_k}=N,y^i_{t_kN}=j,d^i_{t_kNj}=l|\nonumber\\
&S(t_k)=\hat{s}_k,E(t_k)=\hat{e}_k,X(t_{k-1})=\hat{x}(t_{k-1});p,\lambda)\}\times \nonumber\\
&\times P(y^i_{t_kN}=j|f^i_{t_k}=N,S(t_k)=\hat{s}_k,E(t_k)=\hat{e}_k,X(t_{k-1})=\hat{x}(t_{k-1});p,\lambda)\times\nonumber\\
&\times P(f^i_{t_k}=N|S(t_k)=\hat{s}_k,E(t_k)=\hat{e}_k,X(t_{k-1})=\hat{x}(t_{k-1});p,\lambda)- \chi_1-\chi_2.\label{chp3.sec3.1.L10}
\end{align}
Applying similar algebraic manipulations and Jensen's inequality on (\ref{chp3.sec3.1.L10}) we obtain the following,
\begin{align}
&l(p,\lambda|\hat{H}_T)\ge \sum_{k=1}^{T}\sum_{N=0}^{\infty}\sum_{j=0}^{N}\sum_{l=1}^{j}\big[P(f^i_{t_k}=N|X(t_{k-1})=\hat{x}(t_{k-1});p,\lambda)\times \nonumber\\
&\times P(y^i_{t_kN}=j|f^i_{t_k}=N,X(t_{k-1})=\hat{x}(t_{k-1});p,\lambda)\times\nonumber\\
&\times P(d^i_{t_kNj}=l|f^i_{t_k}=N,y^i_{t_kN}=j,X(t_{k-1})=\hat{x}(t_{k-1});p,\lambda)\times\nonumber\\
&\times log\{P(S(t_k)=\hat{s}_k,f^i_{t_k}=N,y^i_{t_kN}=j,d^i_{t_kNj}=l|X(t_{k-1})=\hat{x}(t_{k-1});p,\lambda)\}\big]\nonumber\\
&+ \sum_{k=1}^{T}\sum_{N=0}^{\infty}\sum_{j=0}^{N}\sum_{l=1}^{j}\big[P(f^i_{t_k}=N|S(t_k)=\hat{s}_k,X(t_{k-1})=\hat{x}(t_{k-1});p,\lambda)\times\nonumber\\
&\times P(y^i_{t_kN}=j|f^i_{t_k}=N,S(t_k)=\hat{s}_k,X(t_{k-1})=\hat{x}(t_{k-1});p,\lambda)\times \nonumber\\
&\times P(d^i_{t_kNj}=l|f^i_{t_k}=N,y^i_{t_kN}=j,S(t_k)=\hat{s}_k,X(t_{k-1})=\hat{x}(t_{k-1});p,\lambda)\times \nonumber\\
&\times log\{P(E(t_k)=\hat{e}_k,f^i_{t_k}=N,y^i_{t_kN}=j,d^i_{t_kNj}=l|S(t_k)=\hat{s}_k,X(t_{k-1})=\hat{x}(t_{k-1});p,\lambda)\}\big] \nonumber\\
&+ \sum_{k=1}^{T}\sum_{N=0}^{\infty}\sum_{j=0}^{N}\sum_{l=1}^{j}\big[P(f^i_{t_k}=N|S(t_k)=\hat{s}_k,E(t_k)=\hat{e}_k,X(t_{k-1})=\hat{x}(t_{k-1});p,\lambda)\times\nonumber\\
&\times P(y^i_{t_kN}=j|f^i_{t_k}=N,S(t_k)=\hat{s}_k,E(t_k)=\hat{e}_k,X(t_{k-1})=\hat{x}(t_{k-1});p,\lambda)\times\nonumber\\
&\times P(d^i_{t_kNj}=l|f^i_{t_k}=N,y^i_{t_kN}=j,S(t_k)=\hat{s}_k,E(t_k)=\hat{e}_k,X(t_{k-1})=\hat{x}(t_{k-1});p,\lambda)\times \nonumber\\
&\times log\{P(I(t_k)=\hat{i}_k,f^i_{t_k}=N,y^i_{t_kN}=j,d^i_{t_kNj}=l|S(t_k)=\hat{s}_k,E(t_k)=\hat{e}_k,\nonumber\\
&X(t_{k-1})=\hat{x}(t_{k-1});p,\lambda)\}\big]- \chi_1-\chi_2-\chi_3,\label{chp3.sec3.1.L11}
\end{align}
where,
\begin{align}
&\chi_3 = \sum_{k=1}^{T}\sum_{N=0}^{\infty}\sum_{j=0}^{N}\sum_{l=1}^{j}log\{P(d^i_{t_kNj}=l|f^i_{t_k}=N,y^i_{t_kN}=j,X(t_{k-1})=\hat{x}(t_{k-1});p,\lambda)\}\times \nonumber\\
&\times P(d^i_{t_kNj}=l|f^i_{t_k}=N,y^i_{t_kN}=j,X(t_{k-1})=\hat{x}(t_{k-1});p,\lambda)\nonumber\\
&\times P(y^i_{t_kN}=j|f^i_{t_k}=N,X(t_{k-1})=\hat{x}(t_{k-1});p,\lambda)\times P(f^i_{t_k}=N|X(t_{k-1})=\hat{x}(t_{k-1});p,\lambda)\nonumber\\
&+ \sum_{k=1}^{T}\sum_{N=0}^{\infty}\sum_{j=0}^{N}\sum_{l=1}^{j}log\{P(d^i_{t_kNj}=l|f^i_{t_k}=N,y^i_{t_kN}=j,S(t_k)=\hat{s}_k,X(t_{k-1})=\hat{x}(t_{k-1});\nonumber\\
&p,\lambda)\}\times P(d^i_{t_kNj}=l|f^i_{t_k}=N,y^i_{t_kN}=j,S(t_k)=\hat{s}_k,X(t_{k-1})=\hat{x}(t_{k-1});p,\lambda)\times\nonumber\\
&\times P(y^i_{t_kN}=j|f^i_{t_k}=N,S(t_k)=\hat{s}_k,X(t_{k-1})=\hat{x}(t_{k-1});p,\lambda)\nonumber\\
&\times P(f^i_{t_k}=N|S(t_k)=\hat{s}_k,X(t_{k-1})=\hat{x}(t_{k-1});p,\lambda)\nonumber\\
&+ \sum_{k=1}^{T}\sum_{N=0}^{\infty}\sum_{j=0}^{N}\sum_{l=1}^{j}log\{P(d^i_{t_kNj}=l|f^i_{t_k}=N,y^i_{t_kN}=j,S(t_k)=\hat{s}_k,E(t_k)=\hat{e}_k,\nonumber\\
&X(t_{k-1})=\hat{x}(t_{k-1});p,\lambda)\}
\times P(d^i_{t_kNj}=l|f^i_{t_k}=N,y^i_{t_kN}=j,S(t_k)=\hat{s}_k,E(t_k)=\hat{e}_k,\nonumber\\&X(t_{k-1})=\hat{x}(t_{k-1});p,\lambda)\times P(y^i_{t_kN}=j|f^i_{t_k}=N,S(t_k)=\hat{s}_k,E(t_k)=\hat{e}_k,\nonumber\\
&X(t_{k-1})=\hat{x}(t_{k-1});p,\lambda)\times P(f^i_{t_k}=N|S(t_k)=\hat{s}_k,E(t_k)=\hat{e}_k,X(t_{k-1})=\hat{x}(t_{k-1});p,\lambda)
\end{align}
\end{proof}
\begin{remark}\label{chp3.sec.3.1.1.remark3}
We note from (\ref{chp3.sec3.1.L0}) that the E-step of the EM algorithm consists of finding the conditional expectation term
\begin{align}
&Q(\Theta|\hat{\Theta}^m)= \sum_{k=1}^{T}\sum_{N=0}^{\infty}\sum_{j=0}^{N}\sum_{l=1}^{j}\big[P(f^i_{t_k}=N|X(t_{k-1})=\hat{x}(t_{k-1});\hat{p}^{(m)},\hat{\lambda}^{(m)})\times \nonumber\\
&\times P(y^i_{t_kN}=j|f^i_{t_k}=N,X(t_{k-1})=\hat{x}(t_{k-1});\hat{p}^{(m)},\hat{\lambda}^{(m)})\times\nonumber\\
&\times P(d^i_{t_kNj}=l|f^i_{t_k}=N,y^i_{t_kN}=j,X(t_{k-1})=\hat{x}(t_{k-1});\hat{p}^{(m)},\hat{\lambda}^{(m)})\times\nonumber\\
&\times log\{P(S(t_k)=\hat{s}_k,f^i_{t_k}=N,y^i_{t_kN}=j,d^i_{t_kNj}=l|X(t_{k-1})=\hat{x}(t_{k-1});p,\lambda)\}\big]\nonumber\\
&+ \sum_{k=1}^{T}\sum_{N=0}^{\infty}\sum_{j=0}^{N}\sum_{l=1}^{j}\big[P(f^i_{t_k}=N|S(t_k)=\hat{s}_k,X(t_{k-1})=\hat{x}(t_{k-1});\hat{p}^{(m)},\hat{\lambda}^{(m)})\times\nonumber\\
&\times P(y^i_{t_kN}=j|f^i_{t_k}=N,S(t_k)=\hat{s}_k,X(t_{k-1})=\hat{x}(t_{k-1});\hat{p}^{(m)},\hat{\lambda}^{(m)})\times \nonumber\\
&\times P(d^i_{t_kNj}=l|f^i_{t_k}=N,y^i_{t_kN}=j,S(t_k)=\hat{s}_k,X(t_{k-1})=\hat{x}(t_{k-1});\hat{p}^{(m)},\hat{\lambda}^{(m)})\times \nonumber\\
&\times log\{P(E(t_k)=\hat{e}_k,f^i_{t_k}=N,y^i_{t_kN}=j,d^i_{t_kNj}=l|S(t_k)=\hat{s}_k,X(t_{k-1})=\hat{x}(t_{k-1});p,\lambda)\}\big] \nonumber\\
&+ \sum_{k=1}^{T}\sum_{N=0}^{\infty}\sum_{j=0}^{N}\sum_{l=1}^{j}\big[P(f^i_{t_k}=N|S(t_k)=\hat{s}_k,E(t_k)=\hat{e}_k,X(t_{k-1})=\hat{x}(t_{k-1});\hat{p}^{(m)},\hat{\lambda}^{(m)})\times\nonumber\\
&\times P(y^i_{t_kN}=j|f^i_{t_k}=N,S(t_k)=\hat{s}_k,E(t_k)=\hat{e}_k,X(t_{k-1})=\hat{x}(t_{k-1});\hat{p}^{(m)},\hat{\lambda}^{(m)})\times\nonumber\\
&\times P(d^i_{t_kNj}=l|f^i_{t_k}=N,y^i_{t_kN}=j,S(t_k)=\hat{s}_k,E(t_k)=\hat{e}_k,X(t_{k-1})=\hat{x}(t_{k-1});\hat{p}^{(m)},\hat{\lambda}^{(m)}) \nonumber\\
&\times log\{P(I(t_k)=\hat{i}_k,f^i_{t_k}=N,y^i_{t_kN}=j,d^i_{t_kNj}=l|S(t_k)=\hat{s}_k,E(t_k)=\hat{e}_k,\nonumber\\
&X(t_{k-1})=\hat{x}(t_{k-1});p,\lambda)\}\big],\label{chp3.sec3.1q}
\end{align}
where $\Theta=(p,\lambda)$, and $\Theta^{m}=(\hat{p}^m,\hat{\lambda}^m)$ is the estimate of $(p,\lambda)$ in the $m^{th}$ step of the EM algorithm.
\end{remark}
We specify an explicit expression for components of the E-step Q-function (\ref{chp3.sec3.1q}) in the following result.
\begin{lemma}\label{chp4.lemaforq}
Suppose the conditions of Assumption~\ref{chp2.sec2.assum2} are satisfied, and let $T_1 = T_2 = \Delta t$, and $B(t_k) = D(t_k) = 0$. For each $k \in \{1,2,\ldots T\}, N \ge 0, j \in \{0,1,2,\ldots,N\}$, and $l \in \{1,2,3,\ldots,j\}$, the following hold:
\begin{align}
&P(f^i_{t_k}=N|X(t_{k-1})=\hat{x}(t_{k-1});{\hat{p}}^{(m)},{\hat{\lambda}}^{(m)})=\frac{e^{-{\hat{\lambda}}^{(m)}}(\hat{\lambda}^{(m)})^N}{N!},\label{chp3.sec.3.1.1.lemp9.1}\\
&P(f^i_{t_k}=N|S(t_k)=\hat{s}_k, X(t_{k-1})=\hat{x}(t_{k-1});{\hat{p}}^{(m)},{\hat{\lambda}}^{(m)})=\frac{e^{-{\hat{\lambda}}^{(m)}}(\hat{\lambda}^{(m)})^N}{N!},\label{chp3.sec.3.1.1.lemp9.2}\\
\quad and \quad\nonumber\\
&P(f^i_{t_k}=N|S(t_k)=\hat{s}_k,E(t_k)=\hat{e}_k, X(t_{k-1})=\hat{x}(t_{k-1});{\hat{p}}^{(m)},{\hat{\lambda}}^{(m)})\nonumber\\
&=\frac{e^{-{\hat{\lambda}}^{(m)}}(\hat{\lambda}^{(m)})^N}{N!}.\label{chp3.sec.3.1.1.lemp9.3}
\end{align}
Also,
\begin{align}
&P(y^i_{t_kN}=j|f^i_{t_k}=N,X(t_{k-1})=\hat{x}(t_{k-1});{\hat{p}}^{(m)},{\hat{\lambda}}^{(m)})\nonumber\\
&=\binom{N}{j}\Big(\frac{\hat{I}(t_{k-1})}{\hat{N}(t_{k-1})-1}\Big)^j \Big(1-\frac{\hat{I}(t_{k-1})}{\hat{N}(t_{k-1})-1}\Big)^{N-j},\label{chp3.sec.3.1.1.lemp9.4}\\
&P(y^i_{t_kN}=j|f^i_{t_k}=N,S(t_k)=\hat{s}_k,X(t_{k-1})=\hat{x}(t_{k-1});{\hat{p}}^{(m)},{\hat{\lambda}}^{(m)})\nonumber\\
&=\binom{N}{j}\Big(\frac{\hat{I}(t_{k-1})}{\hat{N}(t_{k-1})-1}\Big)^j \Big(1-\frac{\hat{I}(t_{k-1})}{\hat{N}(t_{k-1})-1}\Big)^{N-j}\label{chp3.sec.3.1.1.lemp9.5},\\
\quad and \quad\nonumber\\
&P(y^i_{t_kN}=j|f^i_{t_k}=N,S(t_k)=\hat{s}_k,E(t_k)=\hat{e}_k,X(t_{k-1})=\hat{x}(t_{k-1});{\hat{p}}^{(m)},{\hat{\lambda}}^{(m)}),\nonumber\\
&=\binom{N}{j}\Big(\frac{\hat{I}(t_{k-1})}{\hat{N}(t_{k-1})-1}\Big)^j \Big(1-\frac{\hat{I}(t_{k-1})}{\hat{N}(t_{k-1})-1}\Big)^{N-j}\label{chp3.sec.3.1.1.lemp9.6}.
\end{align}
Again,
\begin{align}
&P(d^i_{t_kNj}=l|f^i_{t_k}=N,y^i_{t_kN}=j,X(t_{k-1})=\hat{x}(t_{k-1});{\hat{p}}^{(m)},{\hat{\lambda}}^{(m)})= {\hat{p}}^{(m)},\label{chp3.sec.3.1.1.lemp9.7}\\
&P(d^i_{t_kNj}=l|f^i_{t_k}=N,y^i_{t_kN}=j,S(t_k)=\hat{s}_k,X(t_{k-1})=\hat{x}(t_{k-1});{\hat{p}}^{(m)},{\hat{\lambda}}^{(m)})\nonumber\\
&= {\hat{p}}^{(m)},\label{chp3.sec.3.1.1.lemp9.8}\\
\quad and \quad\nonumber\\
&P(d^i_{t_kNj}=l|f^i_{t_k}=N,y^i_{t_kN}=j,S(t_k)=\hat{s}_k,E(t_k)=\hat{e}_k,\nonumber\\
&X(t_{k-1})=\hat{x}(t_{k-1});{\hat{p}}^{(m)},{\hat{\lambda}}^{(m)})= {\hat{p}}^{(m)}.\label{chp3.sec.3.1.1.lemp9.9}
\end{align}
Furthermore,
\begin{align}
&P(S(t_k)=\hat{s}_k,f^i_{t_k}=N,y^i_{t_kN}=j,d^i_{t_kNj}=l|X(t_{k-1})=\hat{x}(t_{k-1});p,\lambda)\nonumber\\
&=\binom{\hat{s}_{k-1}}{\hat{s}_{k-1}-\hat{s}_k}\binom{N}{j}\Big(\frac{\hat{I}(t_{k-1})}{\hat{N}(t_{k-1}))-1}\Big)^j \Big(1-\frac{\hat{I}(t_{k-1})}{\hat{N}(t_{k-1})-1}\Big)^{N-j}\times \nonumber\\
&\times p^{\hat{s}_{k-1}-\hat{s}_k} (1-p)^{\hat{s}_k} p \frac{e^{-\lambda} {\lambda}^N}{N!},\label{chp3.sec.3.1.1.lemp9.10}\\
&P(E(t_k)=\hat{e}_k,f^i_{t_k}=N,y^i_{t_kN}=j,d^i_{t_kNj}=l|S(t_k)=\hat{s}_k,X(t_{k-1})=\hat{x}(t_{k-1});p,\lambda)\nonumber\\
&=\binom{N}{j}\Big(\frac{\hat{I}(t_{k-1})}{\hat{N}(t_{k-1})-1}\Big)^j \Big(1-\frac{\hat{I}(t_{k-1})}{\hat{N}(t_{k-1})-1}\Big)^{N-j}p \frac{e^{-\lambda} {\lambda}^N}{N!},\label{chp3.sec.3.1.1.lemp9.11}\\
&P(I(t_k)=\hat{i}_k,f^i_{t_k}=N,y^i_{t_kN}=j,d^i_{t_kNj}=l|S(t_k)=\hat{s}_k,E(t_k)=\hat{e}_k, X(t_{k-1})=\hat{x}(t_{k-1});\nonumber\\
&p,\lambda)=\binom{N}{j}\Big(\frac{\hat{I}(t_{k-1})}{\hat{N}(t_{k-1})-1}\Big)^j \Big(1-\frac{\hat{I}(t_{k-1})}{\hat{N}(t_{k-1})-1}\Big)^{N-j}p \frac{e^{-\lambda} {\lambda}^N}{N!}.\label{chp3.sec.3.1.1.lemp9.12}
\end{align}
\end{lemma}
\begin{proof}
The equations (\ref{chp3.sec.3.1.1.lemp9.1})-(\ref{chp3.sec.3.1.1.lemp9.9}) follow immediately from Assumption \ref{chp2.sec2.assum2}. For (\ref{chp3.sec.3.1.1.lemp9.10}) we apply the multiplication rule first. That is,
\begin{align}\label{chp3.sec3.1.1.stk1}
&P(S(t_k)=\hat{s}_k,f^i_{t_k}=N,y^i_{t_kN}=j,d^i_{t_kNj}=l|X(t_{k-1})=\hat{x}(t_{k-1});p,\lambda)\nonumber\\
&= P(S(t_k)=\hat{s}_k|f^i_{t_k}=N,y^i_{t_kN}=j,d^i_{t_kNj}=l,X(t_{k-1})=\hat{x}(t_{k-1});p,\lambda)\times\nonumber\\
&\times P(d^i_{t_kNj}=l|f^i_{t_k}=N,y^i_{t_kN}=j,X(t_{k-1})=\hat{x}(t_{k-1});p,\lambda)\times\nonumber\\
&\times P(y^i_{t_kN}=j|f^i_{t_k}=N,X(t_{k-1})=\hat{x}(t_{k-1});p,\lambda)\times\nonumber\\
&\times P(f^i_{t_k}=N|X(t_{k-1})=\hat{x}(t_{k-1});p,\lambda).
\end{align}
When $B(t_k)=0$ and $D(t_k)=0$, then from (\ref{chp2.sec.2.1.3.C_{SE}}) we can get,
\begin{align}\label{chp3.sec3.1.1.Cse}
C_{SE}(t_{k-1})=S(t_{k-1})-S(t_k).
\end{align}
Using (\ref{chp3.sec3.1.1.Cse}) and Assumption \ref{chp2.sec2.assum2} we can write,
\begin{align}\label{chp3.sec3.1.1.stk2}
&P(S(t_k)=\hat{s}_k|f^i_{t_k}=N,y^i_{t_kN}=j,d^i_{t_kNj}=l,X(t_{k-1})=\hat{x}(t_{k-1});p,\lambda)\nonumber\\
&=P(\hat{s}_{k-1}-C_{SE}(t_{k-1})=\hat{s}_k|f^i_{t_k}=N,y^i_{t_kN}=j,d^i_{t_kNj}=l,X(t_{k-1})=\hat{x}(t_{k-1});p,\lambda)\nonumber\\
&=P(C_{SE}(t_{k-1})=\hat{s}_{k-1}-\hat{s}_k|f^i_{t_k}=N,y^i_{t_kN}=j,d^i_{t_kNj}=l,X(t_{k-1})=\hat{x}(t_{k-1});p,\lambda)\nonumber\\
&=\binom{\hat{s}_{k-1}}{\hat{s}_{k-1}-\hat{s}_k} p^{\hat{s}_{k-1}-\hat{s}_k} (1-p)^{\hat{s}_k}.
\end{align}
The equation (\ref{chp3.sec3.1.1.stk2}) follows because, given that the $i^{th}$ infectious person passes infection, then $C_{SE}(t_{k-1})$ is binomial with parameters $p$ and $\hat{s}_{k-1}$. Also, the probability that the $l^{th}$ infectious person passes infection at any time $t_k$, given $j$ infectious individuals present at that time is given by,
\begin{align}\label{chp3.sec3.1.1.di}
P(d^i_{t_kNj}=l|f^i_{t_k}=N,y^i_{t_kN}=j,X(t_{k-1})=\hat{x}(t_{k-1});p,\lambda)=p.
\end{align}
The probability that the $i^{th}$ susceptible person meets $j$ infectious people during epoch $k$ given the $i^{th}$ susceptible person meets $N$ people during that epoch is,
\begin{align}\label{chp3.sec3.1.1.yi}
&P(y^i_{t_kN}=j|f^i_{t_k}=N,X(t_{k-1})=\hat{x}(t_{k-1});p,\lambda)\nonumber\\
&= \binom{N}{j}\Big(\frac{\hat{I}(t_{k-1})}{\hat{N}(t_{k-1})-1}\Big)^j \Big(1-\frac{\hat{I}(t_{k-1})}{\hat{N}(t_{k-1})-1}\Big)^{N-j}.
\end{align}
Also, the probability of the number of people the $i^{th}$ susceptible person meets during the epoch $k$ is
\begin{align}\label{chp3.sec3.1.1.fi}
P(f^i_{t_k}=N|X(t_{k-1})=\hat{x}(t_{k-1});p,\lambda)= \frac{e^{-\lambda} {\lambda}^N}{N!}.
\end{align}
Substituting (\ref{chp3.sec3.1.1.stk2})-(\ref{chp3.sec3.1.1.fi}) into (\ref{chp3.sec3.1.1.stk1}) gives (\ref{chp3.sec.3.1.1.lemp9.10}).
Similarly,
\begin{align}\label{chp3.sec3.1.1.ei}
&P(E(t_k)=\hat{e}_k,f^i_{t_k}=N,y^i_{t_kN}=j,d^i_{t_kNj}=l|S(t_k)=\hat{s}_k,X(t_{k-1})=\hat{x}(t_{k-1});p,\lambda)\nonumber\\
&=P(E(t_k)=\hat{e}_k|f^i_{t_k}=N,y^i_{t_kN}=j,d^i_{t_kNj}=l,S(t_k)=\hat{s}_k,X(t_{k-1})=\hat{x}(t_{k-1});p,\lambda)\times\nonumber\\
&\times P(d^i_{t_kNj}=l|f^i_{t_k}=N,y^i_{t_kN}=j,S(t_k)=\hat{s}_k,X(t_{k-1})=\hat{x}(t_{k-1});p,\lambda)\times\nonumber\\
&\times P(y^i_{t_kN}=j|f^i_{t_k}=N,S(t_k)=\hat{s}_k,X(t_{k-1})=\hat{x}(t_{k-1});p,\lambda)\times\nonumber\\
&\times P(f^i_{t_k}=N|S(t_k)=\hat{s}_k,X(t_{k-1})=\hat{x}(t_{k-1});p,\lambda).
\end{align}
When $B(t_k)=0$ and $D(t_k)=0$, then from (\ref{chp2.sec.2.1.3.C_{EI}}) we can get,
\begin{align}\label{chp3.sec3.1.1.Cei}
&C_{EI}(t_{k-1})=E(t_{k-1})-E(t_k)+S(t_{k-1})-S(t_k)\nonumber\\
\end{align}
Using (\ref{chp3.sec3.1.1.Cei}) we can write,
\begin{align}\label{chp3.sec3.1.1.ei1}
&P(E(t_k)=\hat{e}_k|f^i_{t_k}=N,y^i_{t_kN}=j,d^i_{t_kNj}=l,S(t_k)=\hat{s}_k,X(t_{k-1})=\hat{x}(t_{k-1});p,\lambda)\nonumber\\
&= P(C_{EI}(t_{k-1})=\hat{e}_{k-1}-\hat{e}_k+\hat{s}_{k-1}-\hat{s}_k|f^i_{t_k}=N,y^i_{t_kN}=j,d^i_{t_kNj}=l,S(t_k)=\hat{s}_k,\nonumber\\
&X(t_{k-1})=\hat{x}(t_{k-1});p,\lambda).
\end{align}
Since incubation period $T_1$ is fixed for every person, and equal to $\Delta t$, then all exposed person at the beginning of epoch $k$ become infectious at the beginning of epoch $k+1$. It is easy to see that,
\begin{align}\label{chp3.sec3.1.1.ei2}
C_{EI}(t_{k-1})=E(t_{k-1})=\hat{e}_{k-1}.
\end{align}
So (\ref{chp3.sec3.1.1.ei1}) can be rewritten as,
\begin{align}\label{chp3.sec3.1.1.ei3}
&P(E(t_k)=\hat{e}_k|f^i_{t_k}=N,y^i_{t_kN}=j,d^i_{t_kNj}=l,S(t_k)=\hat{s}_k,X(t_{k-1})=\hat{x}(t_{k-1});p,\lambda)\nonumber\\
&=\begin{cases}
  1, & \text{where, $\hat{e}_k=\hat{s}_{k-1}-\hat{s}_k$},\\
  0, &\text{otherwise.}
  \end{cases}
\end{align}
Moreover, all the other components of (\ref{chp3.sec3.1.1.ei}) are obtained similarly as in (\ref{chp3.sec3.1.1.di})-(\ref{chp3.sec3.1.1.fi}). Substituting the obtained components and (\ref{chp3.sec3.1.1.ei3}) into (\ref{chp3.sec3.1.1.ei}) gives (\ref{chp3.sec.3.1.1.lemp9.11}).\par
Furthermore, we can write
\begin{align}\label{chp3.sec3.1.1.i1}
&P(I(t_k)=\hat{i}_k,f^i_{t_k}=N,y^i_{t_kN}=j,d^i_{t_kNj}=l|S(t_k)=\hat{s}_k,E(t_k)=\hat{e}_k,\nonumber\\
&X(t_{k-1})=\hat{x}(t_{k-1});p,\lambda)=P(I(t_k)=\hat{i}_k|f^i_{t_k}=N,y^i_{t_kN}=j,d^i_{t_kNj}=l,S(t_k)=\hat{s}_k,\nonumber\\
&E(t_k)=\hat{e}_k,X(t_{k-1})=\hat{x}(t_{k-1});p,\lambda)\times P(d^i_{t_kNj}=l|f^i_{t_k}=N,y^i_{t_kN}=j,S(t_k)=\hat{s}_k,\nonumber\\
&E(t_k)=\hat{e}_k,X(t_{k-1})=\hat{x}(t_{k-1});p,\lambda)\times\nonumber\\
&\times P(y^i_{t_kN}=j|f^i_{t_k}=N,S(t_k)=\hat{s}_k,E(t_k)=\hat{e}_k,X(t_{k-1})=\hat{x}(t_{k-1});p,\lambda)\times\nonumber\\
&\times P(f^i_{t_k}=N|S(t_k)=\hat{s}_k,E(t_k)=\hat{e}_k,X(t_{k-1})=\hat{x}(t_{k-1});p,\lambda).
\end{align}
When $B(t_k)=0$ and $D(t_k)=0$, then from (\ref{chp2.sec2.1.3.C_{IR}}) we can get,
\begin{align}\label{chp3.sec3.1.1.Cir}
&C_{IR}(t_{k-1})=I(t_{k-1})-I(t_k)+ E(t_{k-1})-E(t_k)+S(t_{k-1})-S(t_k),\nonumber\\
\end{align}
Using (\ref{chp3.sec3.1.1.Cir}), we can write
\begin{align}\label{chp3.sec3.1.1.ir1}
&P(I(t_k)=\hat{i}_k|f^i_{t_k}=N,y^i_{t_kN}=j,d^i_{t_kNj}=l,S(t_k)=\hat{s}_k,E(t_k)=\hat{e}_k,X(t_{k-1})=\hat{x}(t_{k-1});\nonumber\\
&p,\lambda)=P(\hat{i}_{k-1}-C_{IR}(t_{k-1})+\hat{e}_{k-1}-\hat{e}_k+\hat{s}_{k-1}-\hat{s}_k=\hat{i}_k|f^i_{t_k}=N,y^i_{t_kN}=j,d^i_{t_kNj}=l,\nonumber\\
&S(t_k)=\hat{s}_k,E(t_k)=\hat{e}_k,X(t_{k-1})=\hat{x}(t_{k-1});p,\lambda)\nonumber\\
&=P(C_{IR}(t_{k-1})=\hat{i}_{k-1}-\hat{i}_k+\hat{e}_{k-1}-\hat{e}_k+\hat{s}_{k-1}-\hat{s}_k|f^i_{t_k}=N,y^i_{t_kN}=j,d^i_{t_kNj}=l,\nonumber\\
&S(t_k)=\hat{s}_k,E(t_k)=\hat{e}_k,X(t_{k-1})=\hat{x}(t_{k-1});p,\lambda).
\end{align}
Since, infectious period $T_2$ is fixed for every person, and equal to one unit time $\Delta t$, then all infectious persons at the beginning of epoch $k$ become infectious at the beginning of epoch $k+1$. It is easy to see that,
\begin{align}\label{chp3.sec3.1.1.ir2}
C_{IR}(t_{k-1})=I(t_{k-1})=\hat{i}_{k-1}.
\end{align}
So, (\ref{chp3.sec3.1.1.ir1}) can be rewritten as
\begin{align}\label{chp3.sec3.1.1.ir3}
&P(I(t_k)=\hat{i}_k|f^i_{t_k}=N,y^i_{t_kN}=j,d^i_{t_kNj}=l,S(t_k)=\hat{s}_k,E(t_k)=\hat{e}_k,\nonumber\\
& X(t_{k-1})=\hat{x}(t_{k-1});p,\lambda)\nonumber\\
&= \begin{cases}
  1, & \text{where, $\hat{i}_k=\hat{e}_{k-1}-\hat{e}_k+\hat{s}_{k-1}-\hat{s}_k$},\\
  0, &\text{otherwise.}
  \end{cases}
\end{align}
Moreover, all the other components of (\ref{chp3.sec3.1.1.i1}) are obtained similarly as in (\ref{chp3.sec3.1.1.di})-(\ref{chp3.sec3.1.1.fi}). Substituting the obtained components and (\ref{chp3.sec3.1.1.ir3}) into (\ref{chp3.sec3.1.1.i1}) gives (\ref{chp3.sec.3.1.1.lemp9.12}).
\end{proof}
The following result presents an explicit expression for the E-step Q-function of the EM algorithm.
\begin{theorem}\label{chp3.sec3.1.1thm1}
For $m=0,1,2,\ldots$, the E-step Q-function of the EM algorithm in (\ref{chp3.sec3.1q}) in Remark \ref{chp3.sec.3.1.1.remark3} is expressed as follows for $\Theta=(p,\lambda)$
\begin{align}
Q(\Theta|\hat{\Theta}^{(m)}) &\equiv Q(p,\lambda|\hat{p}^{(m)},\hat{\lambda}^{(m)})\nonumber\\&= \Re + \sum_{k=1}^{T}\sum_{N=0}^{\infty}\frac{e^{-{\hat{\lambda}}^{(m)}}(\hat{\lambda}^{(m)})^N}{N!}\times N\Big(\frac{\hat{I}(t_{k-1})}{\hat{N}(t_{k-1})-1}\Big)\times\hat{p}^{(m)}\times\nonumber\\
&\times \Big[(\hat{s}_{k-1}-\hat{s}_k)\log(p) + (\hat{s}_k)\log(1-p)+ \log(p)-\lambda \nonumber\\
&+ N \log(\lambda)+ \log(p)-\lambda + N \log(\lambda)+\log(p)-\lambda + N \log(\lambda) \Big],\nonumber\\
&= \Re + \sum_{k=1}^{T}\sum_{N=0}^{\infty}\frac{e^{-{\hat{\lambda}}^{(m)}}(\hat{\lambda}^{(m)})^N}{N!}\times N\Big(\frac{\hat{I}(t_{k-1})}{\hat{N}(t_{k-1})-1}\Big)\times\hat{p}^{(m)}\times\nonumber\\
&\times \Big[(\hat{s}_{k-1}-\hat{s}_k)\log(p) + (\hat{s}_k)\log(1-p)+ 3\log(p)+3N\log(\lambda)-3\lambda\Big],\label{chp3.sec3.1.1.estep}
\end{align}
where $\Re$ is a constant term that dependents only on the estimates $\Theta^{m}=(p^{m},\lambda^{m})$.
\end{theorem}
\begin{proof}
From Lemma \ref{chp4.lemaforq} and (\ref{chp3.sec3.1q}), it is easy to see that
\begin{align}\label{chp3.sec3.1.1.theta_estimated}
Q(\Theta|\hat{\Theta}^{(m)}) &\equiv Q(p,\lambda|\hat{p}^{(m)},\hat{\lambda}^{(m)})\nonumber\\&= \sum_{k=1}^{T}\sum_{N=0}^{\infty}\sum_{j=0}^{N}\sum_{l=1}^{j}\frac{e^{-{\hat{\lambda}}^{(m)}}(\hat{\lambda}^{(m)})^N}{N!}\times \binom{N}{j}\Big(\frac{\hat{I}(t_{k-1})}{\hat{N}(t_{k-1})-1}\Big)^j \nonumber\\
&\Big(1-\frac{\hat{I}(t_{k-1})}{\hat{N}(t_{k-1})-1}\Big)^{N-j} \hat{p}^{(m)} \times\nonumber\\
&\times \Big\{\log\Big[\binom{\hat{s}_{k-1}}{\hat{s}_{k-1}-\hat{s}_k}\binom{N}{j}\Big(\frac{\hat{I}(t_{k-1})}{\hat{N}(t_{k-1})-1}\Big)^j \Big(1-\frac{\hat{I}(t_{k-1})}{\hat{N}(t_{k-1})-1}\Big)^{N-j}\Big]\nonumber\\
&-\log(N!)+(\hat{s}_{k-1}-\hat{s}_k)\log(p) + (\hat{s}_k)\log(1-p)+ \log(p)-\lambda + N \log(\lambda)\Big\}\nonumber\\
&+ \sum_{k=1}^{T}\sum_{N=0}^{\infty}\sum_{j=0}^{N}\sum_{l=1}^{j}\frac{e^{-{\hat{\lambda}}^{(m)}}(\hat{\lambda}^{(m)})^N}{N!}\times \binom{N}{j}\Big(\frac{\hat{I}(t_{k-1})}{\hat{N}(t_{k-1})-1}\Big)^j\nonumber\\
& \Big(1-\frac{\hat{I}(t_{k-1})}{\hat{N}(t_{k-1})-1}\Big)^{N-j} \hat{p}^{(m)} \times\nonumber\\
&\times \Big\{\log\Big[\binom{N}{j}\Big(\frac{\hat{I}(t_{k-1})}{\hat{N}(t_{k-1})-1}\Big)^j \Big(1-\frac{\hat{I}(t_{k-1})}{\hat{N}(t_{k-1})-1}\Big)^{N-j}\Big]\nonumber\\
&-\log(N!)+ \log(p)-\lambda + N \log(\lambda)\Big\}\nonumber\\
&+ \sum_{k=1}^{T}\sum_{N=0}^{\infty}\sum_{j=0}^{N}\sum_{l=1}^{j}\frac{e^{-{\hat{\lambda}}^{(m)}}(\hat{\lambda}^{(m)})^N}{N!}\times \binom{N}{j}\Big(\frac{\hat{I}(t_{k-1})}{\hat{N}(t_{k-1})-1}\Big)^j\nonumber\\
& \Big(1-\frac{\hat{I}(t_{k-1})}{\hat{N}(t_{k-1})-1}\Big)^{N-j} \hat{p}^{(m)} \times\nonumber\\
&\times \Big\{\log\Big[\binom{N}{j}\Big(\frac{\hat{I}(t_{k-1})}{\hat{N}(t_{k-1})-1}\Big)^j \Big(1-\frac{\hat{I}(t_{k-1})}{\hat{N}(t_{k-1})-1}\Big)^{N-j}\Big]-\log(N!)+\nonumber\\
&+ \log(p)-\lambda + N \log(\lambda)\Big\}.\nonumber\\
\end{align}
Observe that
\begin{align}
\sum_{j=0}^{N}\sum_{l=1}^{j}\binom{N}{j}\Big(\frac{\hat{I}(t_{k-1})}{\hat{N}(t_{k-1})-1}\Big)^j \Big(1-\frac{\hat{I}(t_{k-1})}{\hat{N}(t_{k-1})-1}\Big)^{N-j}= N\Big(\frac{\hat{I}(t_{k-1})}{\hat{N}(t_{k-1})-1}\Big).
\end{align}
Thus, (\ref{chp3.sec3.1.1.estep}) follows immediately from (\ref{chp3.sec3.1.1.theta_estimated}), where
\begin{align}\label{chp3.sec3.1.1.theta_estimatedre}
\Re &\equiv  \sum_{k=1}^{T}\sum_{N=0}^{\infty}\sum_{j=0}^{N}\sum_{l=1}^{j}\frac{e^{-{\hat{\lambda}}^{(m)}}(\hat{\lambda}^{(m)})^N}{N!}\times \binom{N}{j}\Big(\frac{\hat{I}(t_{k-1})}{\hat{N}(t_{k-1})-1}\Big)^j \Big(1-\frac{\hat{I}(t_{k-1})}{\hat{N}(t_{k-1})-1}\Big)^{N-j} \hat{p}^{(m)} \nonumber\\
&\times \Big\{\log\Big[\binom{\hat{s}_{k-1}}{\hat{s}_{k-1}-\hat{s}_k}\binom{N}{j}\Big(\frac{\hat{I}(t_{k-1})}{\hat{N}(t_{k-1})-1}\Big)^j \Big(1-\frac{\hat{I}(t_{k-1})}{\hat{N}(t_{k-1})-1}\Big)^{N-j}\Big]-\log(N!)\Big\}\nonumber\\
&+ \sum_{k=1}^{T}\sum_{N=0}^{\infty}\sum_{j=0}^{N}\sum_{l=1}^{j}\frac{e^{-{\hat{\lambda}}^{(m)}}(\hat{\lambda}^{(m)})^N}{N!}\times \binom{N}{j}\Big(\frac{\hat{I}(t_{k-1})}{\hat{N}(t_{k-1})-1}\Big)^j \Big(1-\frac{\hat{I}(t_{k-1})}{\hat{N}(t_{k-1})-1}\Big)^{N-j} \hat{p}^{(m)} \nonumber\\
&\times \Big\{\log \Big[\binom{N}{j}\Big(\frac{\hat{I}(t_{k-1})}{\hat{N}(t_{k-1})-1}\Big)^j \Big(1-\frac{\hat{I}(t_{k-1})}{\hat{N}(t_{k-1})-1}\Big)^{N-j}\Big]-\log(N!)\Big\}\nonumber\\
&+ \sum_{k=1}^{T}\sum_{N=0}^{\infty}\sum_{j=0}^{N}\sum_{l=1}^{j}\frac{e^{-{\hat{\lambda}}^{(m)}}(\hat{\lambda}^{(m)})^N}{N!}\times \binom{N}{j}\Big(\frac{\hat{I}(t_{k-1})}{\hat{N}(t_{k-1})-1}\Big)^j \Big(1-\frac{\hat{I}(t_{k-1})}{\hat{N}(t_{k-1})-1}\Big)^{N-j} \hat{p}^{(m)} \nonumber\\
&\times \Big\{\log\Big[\binom{N}{j}\Big(\frac{\hat{I}(t_{k-1})}{\hat{N}(t_{k-1})-1}\Big)^j \Big(1-\frac{\hat{I}(t_{k-1})}{\hat{N}(t_{k-1})-1}\Big)^{N-j}\Big]-\log(N!)\Big\}.
\end{align}
\end{proof}
\begin{remark} \label{chp3.sec3.1.1remark4}
It follows from Theorem \ref{chp3.sec3.1.1thm1} that the M-step of the EM algorithm consists of maximizing $Q(p,\lambda|\hat{p}^{(m)},\hat{\lambda}^{(m)})$ with respect to $p, \lambda$. This is equivalent to maximizing the non-constant term of (\ref{chp3.sec3.1.1.estep}).
\end{remark}\par
In the next result, we present the M-step of the EM algorithm, and an explicit MLE for $p, \lambda$.
\begin{theorem}\label{chp3.sec3.1.1remark4.thm1}
Let the E-step of the EM algorithm be as defined in Theorem \ref{chp3.sec3.1.1thm1}. Then the MLE $\hat{p}$ of $p$ is given as follows:
\begin{align}
\hat{p} &= \frac{\sum_{k=1}^{T}\Big(\frac{\hat{I}(t_{k-1})}{\hat{N}(t_{k-1})-1}\Big)\times (\hat{s}_{k-1}+3)-\sum_{k=1}^{T}\Big(\frac{\hat{I}(t_{k-1})}{\hat{N}(t_{k-1})-1}\Big)\times \hat{s}_k}{\sum_{k=1}^{T}\Big(\frac{\hat{I}(t_{k-1})}{\hat{N}(t_{k-1})-1}\Big)\times (\hat{s}_{k-1}+3)}\label{chp4.thmprof2.1}\\
&= 1 - \frac{\sum_{k=1}^{T}\Big(\frac{\hat{I}(t_{k-1})}{\hat{N}(t_{k-1})-1}\Big)\times \hat{s}_k}{\sum_{k=1}^{T}\Big(\frac{\hat{I}(t_{k-1})}{\hat{N}(t_{k-1})-1}\Big)\times (\hat{s}_{k-1}+3)}\label{chp4.thmprof2.2}
\end{align}
The estimate of the MLE $\hat{\lambda}$ of $\lambda$ at the $m^{th}$ step is given as follows:
\begin{align}
\hat{\lambda}^{(m+1)} &= \frac{\sum_{k=1}^{T}\Big(1 + \hat{\lambda}^{(m)}\Big)\Big(\frac{\hat{I}(t_{k-1})}{\hat{N}(t_{k-1})-1}\Big)}{\sum_{k=1}^{T}\Big(\frac{\hat{I}(t_{k-1})}{\hat{N}(t_{k-1})-1}\Big)},\label{chp4.thmprof2.3}\nonumber\\
& m=0,1,2,\ldots
\end{align}
\end{theorem}
\begin{proof}
From (\ref{chp3.sec3.1.1.estep}), observe that at the $m^{th}$ step, $m=0,1,2,\ldots$, maximizing the E-step Q-function $Q(\Theta|\hat{\Theta}^{(m)})$ with respect to $p$ and $\lambda$, consists of taking the partial derivatives of $Q(\Theta|\hat{\Theta}^{(m)})$ with respect to $p$ and $\lambda$, that is,
\begin{align}
&\frac{\partial Q}{\partial p} = \sum_{k=1}^{T}\sum_{N=0}^{\infty}\frac{e^{-{\hat{\lambda}}^{(m)}}(\hat{\lambda}^{(m)})^N}{N!}\times N\Big(\frac{\hat{I}(t_{k-1})}{\hat{N}(t_{k-1})-1}\Big)\times\nonumber\\
&\times \hat{p}^{(m)}\times\Big[\frac{(\hat{s}_{k-1}-\hat{s}_k)}{p}-\frac{\hat{s}_k}{1-p}+\frac{3}{p}\Big],\label{chp5.sec4.2.deltap1}\\
&\frac{\partial Q}{\partial \lambda} = \sum_{k=1}^{T}\sum_{N=0}^{\infty}\frac{e^{-{\hat{\lambda}}^{(m)}}(\hat{\lambda}^{(m)})^N}{N!}\times N\Big(\frac{\hat{I}(t_{k-1})}{\hat{N}(t_{k-1})-1}\Big)\times\hat{p}^{(m)}\times\Big[\frac{3N}{\lambda}-3\Big]\label{chp5.sec4.2.deltaq1}.
\end{align}

Observe that the terms,
\begin{align}
\sum_{N=0}^{\infty}\frac{e^{-{\hat{\lambda}}^{(m)}}(\hat{\lambda}^{(m)})^N}{N!}\times N = \hat{\lambda}^{(m)},\label{chp5.sec4.2.lamda1}\\
\sum_{N=0}^{\infty}N^2\times \frac{e^{-{\hat{\lambda}}^{(m)}}(\hat{\lambda}^{(m)})^N}{N!}=\hat{\lambda}^{(m)}+(\hat{\lambda}^{(m)})^2.\label{chp5.sec4.2.lamda2}
\end{align}
Using (\ref{chp5.sec4.2.lamda1}) into (\ref{chp5.sec4.2.deltap1}), and setting the result to zero, we get
\begin{align}
\frac{\partial Q}{\partial p}
 &= \sum_{k=1}^{T}\hat{\lambda}^{(m)}\Big(\frac{\hat{I}(t_{k-1})}{\hat{N}(t_{k-1})-1}\Big)\times \hat{p}^{(m)}\times\Big[\frac{(\hat{s}_{k-1}-\hat{s}_k)}{p}-\frac{\hat{s}_k}{1-p}+\frac{3}{p}\Big]=0.\label{chp5.sec4.2p2}
\end{align}
Since $(\hat{p}^{(m)},\hat{\lambda}^{(m)})> 0$, solving for $p$ from (\ref{chp5.sec4.2p2}) leads to (\ref{chp4.thmprof2.2}).

Also, using (\ref{chp5.sec4.2.lamda2}) into (\ref{chp5.sec4.2.deltaq1}), and setting the result to zero, we get
\begin{align}
\frac{\partial Q}{\partial \lambda}= 3 \hat{p}^{(m)} \hat{\lambda}^{(m)}\Big[\sum_{k=1}^{T}(1+\hat{\lambda}^{(m)})\Big(\frac{\hat{I}(t_{k-1})}{\hat{N}(t_{k-1})-1}\Big)\frac{1}{\lambda}-\sum_{k=1}^{T}\Big(\frac{\hat{I}(t_{k-1})}{\hat{N}(t_{k-1})-1}\Big)\Big]=0.\label{chp5.sec4.2lamda2}
\end{align}
Since $(\hat{p}^{(m)},\hat{\lambda}^{(m)})> 0$, solving for $\lambda$ from (\ref{chp5.sec4.2lamda2}) leads to (\ref{chp4.thmprof2.3}).
\end{proof}

\begin{remark}
Observe from (\ref{chp4.thmprof2.1})-(\ref{chp4.thmprof2.2}) that for $\forall k\ge 1, \hat{s_k}\le \hat{s}_{k-1}$, since the population is continuously infected, exposed, and removed. This implies that the numerator of (\ref{chp4.thmprof2.1}) is smaller than the denominator. Further, given the population $X(t_{k})=(S(t_{k}),E(t_{k}),I(t_{k}),R(t_{k}))$  present at time $k\geq 0$, the MLE $\hat{p}$ can be interpreted hypothetically as follows.

 If meeting an infectious person is random, but passing infection is almost sure, then $\alpha^{i}(t_{k}),k\geq 0$ in (\ref{chp2.sec1.alfa}) can also be interpreted as the probability of meeting and almost surely getting infection at time $k\geq 0$. The  term $(\hat{s}_{k-1}+3)$ can be interpreted in one way as representing the critical population present at time $t_{k-1}$, comprising of the susceptible state $\hat{s}_{k-1}$ and one exposed, one infectious and one removed persons. That is, if $\hat{s}_{k-1}=0$ at $t_{k+1}$, there is still possibility of the susceptible person getting infected since $(\hat{s}_{k-1}+3)=3$. This implies that $(\hat{s}_{k-1}+3)$ is the critical population necessary for infection to occur at time $t_{k-1}$.   Thus, $\forall k\ge 1,$ the term $\Big(\frac{\hat{I}(t_{k-1})}{\hat{N}(t_{k-1})-1}\Big)\times (\hat{s}_{k-1}+3)$ is the critical average number of newly infected persons the occur at time $t_{k-1}$ in the critical population of size $(\hat{s}_{k-1}+3)$. It follows that the sum $\sum_{k=1}^{T}\Big(\frac{\hat{I}(t_{k-1})}{\hat{N}(t_{k-1})-1}\Big)\times (\hat{s}_{k-1}+3)$ is the cumulative critical average number of newly infected persons that occur in the observed data $\hat{H}_T$ up to the time $t_{k-1}$.

Since the parameter $p$ is assumed constant in the population at every time step, it is necessary to assume that the infectivity conditions are the same in the next immidiate time $t_{k}$, so that the probability of meeting and getting infection is the same $\Big(\frac{\hat{I}(t_{k-1})}{\hat{N}(t_{k-1})-1}\Big)$. In addition, since the incubation and infectious periods are equal to $\Delta t$, then the critical population at time $t_{k}$ is now $\hat{s}_{k}$, and assuming that infection takes place over the interval $[t_{k-1}, t_k)$, then  $\hat{s_k}< (\hat{s}_{k-1}+3)$. Moreover, the sum $\sum_{k=1}^{T}\Big(\frac{\hat{I}(t_{k-1})}{\hat{N}(t_{k-1})-1}\Big)\times \hat{s}_k$ represents the critical cumulative average number of newly infected people that occur in the observed data up to the time $t_k$.

Therefore, the increment $\sum_{k=1}^{T}\Big(\frac{\hat{I}(t_{k-1})}{\hat{N}(t_{k-1})-1}\Big)\times (\hat{s}_{k-1}+3)-\sum_{k=1}^{T}\Big(\frac{\hat{I}(t_{k-1})}{\hat{N}(t_{k-1})-1}\Big)\times \hat{s}_k$ is the critical average number of newly infected people that occur between the times $t_{k-1}$  and $t_{k}$, that is, over the intervals $[t_{k-1}, t_k)$ and $[t_{k}, t_{k+1})$.
Hence, $\hat{p}$ represents the critical fraction of newly infected people that occur in over a one time unit $\Delta t$. 

Since from (\ref{chp4.thmprof2.3}) the estimate of the MLE $\hat{\lambda}^{(m+1)}$ for $\lambda$ depends on the  $m^{th}$ step estimate $\hat{\lambda}^{(m)}$, we omit the interpretation of the expression in (\ref{chp4.thmprof2.3}). 
\end{remark}
\section{Conclusion}
In this study, we have sufficiently defined a general class of SEIR Markov chain models for infectious diseases such as pneumonia or influenza, which effectively shows the progression of the disease over time for an individual in the population. Moreover, we defined the transition probabilities for the general model.

Furthermore, we presented special SEIR Markov chain models along with their transition probabilities for the disease with (1) zero and nonzero births and deaths, and (2) with fixed or random incubation and infectious periods. We derived the probability that an susceptible person gets infection at time $k$ ($i.e.$ in $[t_k, t_{k+1}$)), and found the conditional distribution of the driving events of the population. In addition, to specify the transition probability of the model for random incubation and infectious periods, we also derived the probabilities that an exposed individual becomes infectious, and the infectious individual becomes recovered at any time interval $[t_k, t_{k+1})$, respectively.

We further applied the expectation maximization (EM) algorithm to find the maximum likelihood estimator of $p$, the probability of passing infection after one interaction with an infectious person at any time, and also for $\lambda$, the average number of people a susceptible individual meets per unit time.

Finally, we presented examples, where we numerically simulate an SEIR epidemic to assess the behavior  of the sample paths for two different cases involving fixed and random incubation and infectious periods of the disease, in order to validate the epidemic models.
\section*{Acknowledgements}
This work was complete during Chinmoy's graduate studies in the Department of Mathematical Sciences, Georgia Southern University in  May 2019. The title of his thesis is "\textit{Studying the stochastic dynamics of pneumonia epidemics: chain-binomial modeling, maximum likelihood estimation and expectation maximization algorithm}". Chinmoy is currently a Ph.D. student in the Department of Mathematics and Statistics, University of Calgary, Canada.


\end{document}